\documentclass[fleqn,usenatbib]{mnras}

\usepackage{newtxtext,newtxmath}
\usepackage[T1]{fontenc}

\DeclareRobustCommand{\VAN}[3]{#2}
\let\VANthebibliography\thebibliography
\def\thebibliography{\DeclareRobustCommand{\VAN}[3]{##3}\VANthebibliography}

\usepackage{graphicx}	% Including figure files
\usepackage{amsmath}
\usepackage{subfigure}% Advanced maths commands
\usepackage{orcidlink}

\newcommand\HI{\hbox{H$\,\rm \scriptstyle I$}~}
\newcommand\HII{\hbox{H$\,\rm \scriptstyle II$}~} 

\newcommand\HeII{\hbox{He$\,\rm \scriptstyle II$}~}
 
\newcommand{\Lya}{Ly-$\alpha$ } 
\newcommand{\angstrom}{\mbox{\normalfont\AA}}

\defcitealias{miralda-escude2000}{MHR00}
\newcommand{\ud}{{\rm d}}
\newcommand{\xhi}{x_{\rm HI}}

\newcommand{\mhr}{\citetalias{miralda-escude2000}}
\newcommand{\eref}[1]{Eq.~\eqref{#1}}
\newcommand{\fref}[1]{Fig.~\ref{#1}}

\title[The Lyman-limit  photon mean free path]{The Lyman-limit  photon mean free path at the end of late reionisation in the Sherwood-Relics simulations}

\author[J. Feron et al.]{Jennifer Feron$^{1}$\,\orcidlink{0009-0001-8626-540X}\thanks{E-mail: ppxjf3@nottingham.ac.uk},
Luke Conaboy$^{1}$\,\orcidlink{0000-0002-6580-7177},
James S. Bolton$^{1}$\,\orcidlink{0000-0003-2764-8248},
Emma Chapman$^{1}$\,\orcidlink{0000-0002-5050-9847},
Martin G. Haehnelt$^{2}$\,\orcidlink{0000-0001-8443-2393}, 
\newauthor Laura C. Keating$^{3}$\,\orcidlink{0000-0001-5211-1958}, 
Girish Kulkarni$^{4}$\,\orcidlink{0000-0001-5829-4716}
and Ewald Puchwein$^{5}$\,\orcidlink{0000-0001-8778-7587}\\
$^{1}$School of Physics and Astronomy, The University of Nottingham, University Park, Nottingham, NG7 2RD, UK\\
$^{2}$Kavli Institute for Cosmology and Institute of Astronomy, Madingley Road, Cambridge, CB3 0HA, UK\\
$^{3}$Institute for Astronomy, University of Edinburgh, Blackford Hill, Edinburgh, EH9 3HJ, UK \\
$^{4}$Tata Institute of Fundamental Research, Homi Bhabha Road, Mumbai 400005, India\\
$^{5}$Leibniz-Institut f\"ur Astrophysik Potsdam, An der Sternwarte 16, 14482 Potsdam, Germany\\}

\date{Accepted XXX. Received YYY; in original form ZZZ}

\pubyear{2024}

\begin{document}
\label{firstpage}
\pagerange{\pageref{firstpage}--\pageref{lastpage}}
\maketitle

\begin{abstract}
Recent evidence supporting reionisation ending at redshift $z \lesssim 6$ includes the rapid redshift evolution of the mean free path, $\lambda_{\rm mfp}$, for Lyman-limit photons through the intergalactic medium at $5<z<6$.  Here we investigate $\lambda_{\rm mfp}$ predicted by the Sherwood-Relics suite of hybrid radiation hydrodynamical simulations.  Simulations with comoving volumes of $40^{3}\,h^{-3}\rm\,cMpc^{3}$ ($160^{3}\,h^{-3}\rm\,cMpc^{3}$), calibrated to match the observed \Lya forest transmission with a late end to reionisation at $z<6$, are consistent with recent $\lambda_{\rm mfp}$ measurements at $z<5.9$, and are $1.2\sigma$ ($1.8\sigma$) above the highest redshift $\lambda_{\rm mfp}$ measurement at $z=5.93$.  The majority of the Lyman-limit opacity at the end of reionisation is attributable to highly ionised \Lya forest absorbers with neutral hydrogen column densities $N_{\rm HI}\leq 10^{16}$--$10^{17}\rm\,cm^{-2}$.  Resolving these systems is critical for capturing the redshift evolution of $\lambda_{\rm mfp}$ during the final stages of reionisation.   After reionisation completes, overdense gas will reduce $\lambda_{\rm mfp}$ by up to 20 per cent around haloes with masses $M_{\rm h}\sim 10^{9}$--$10^{11}h^{-1}\,{\rm M}_{\odot}$, but during reionisation ionised bubbles will instead boost $\lambda_{\rm mfp}$ around haloes by up to an order of magnitude when the IGM is as much as 90 per cent neutral by volume.  This effect will play an important role in the visibility of \Lya emitting galaxies at $z\gtrsim 10$ discovered with JWST.
\end{abstract}


%%%%%%%%%%%%%%%%%%%%%%%%%%%%%%%%%%%%%%%%%%%%%%%%%%%%%%%%%%%%%%%%%%
%%%%%%%%%%%%%%%%%%%%%%%%%%% SECTION 1 %%%%%%%%%%%%%%%%%%%%%%%%%%%%
%%%%%%%%%%%%%%%%%%%%%%%%%%%%%%%%%%%%%%%%%%%%%%%%%%%%%%%%%%%%%%%%%%

\section{Introduction}

There are still many unanswered questions about the epoch of reionisation, including the precise nature of the ionising sources \citep[e.g.][]{Finkelstein_conditions_2019,Naidu2020,Grazian2023,Atek2024}, and the timing and extent of reionisation.  At present, many observations indicate that reionisation has a midpoint around $z \sim 7$--$8$, with the final neutral islands in the diffuse intergalactic medium (IGM) persisting until at least $z\sim 6$ \citep[e.g.][]{Davies2018,Mason2018,Kulkarni2019,Yang_opical_depth_2020,Qin2021,Nakane2023,Umeda_nhi_2023,Durovcikova2024}. 
The last stages of reionisation may even extend as late as $z\sim5.3$, based on the observed fluctuations in the \Lya forest transmission \citep{Becker2015,eilers_opacity_2018,bosman_new_2018,bosman_hydrogen_2022} and the incidence of dark gaps and pixels in the \Lya and Ly-$\beta$ forests \citep{zhu_dark_gaps_2022,jin_nearly_2023}.

Further evidence in support of an end to reionisation at $z\lesssim 6$ is provided by the redshift evolution of the mean free path of Lyman-limit photons through the IGM \cite[e.g.][]{Prochaska_opacity_2009,Fumagalli_hydrogen_2013,OMeara_LLS_2013,Worseck2014,Lusso2018,Romano2019}.  Recent measurements have demonstrated the mean free path declines by an order of magnitude from $z=5$ to $z=6$ \citep{becker_mean_2021,Bosman2021,davies_constraints_2023,gaikwad_measuring_2023,zhu_mfp_2023}.  This evolution is expected if reionisation ends below or near to $z=6$; the mean free path should increase rapidly as ionised hydrogen bubbles merge.  The background photoionisation rate will then increase and the incidence of Lyman-limit photon sinks will decrease \citep[e.g.][]{Wyithe_bias_2008,Park2016,Davies2021,Cain2021,Nasir_photon_sink_2021,Theuns_LLS_2024}.  

Intriguingly, recent numerical simulations of inhomogeneous reionisation have exhibited mixed success in reproducing the observed rapid decline in the mean free path approaching $z\simeq 6$.  Using radiative transfer simulations applied to a hydrodynamical simulation in post-processing, \citet{Keating2020,keating_constraining_2020} found a mean free path evolution that was too gradual, matching the data at $z\sim 5.5$ but overshooting at $z\sim 6$.  The \citet{Keating2020,keating_constraining_2020} model was calibrated to reproduce the large-scale fluctuations observed in the \Lya forest transmission at $z<6$.  In contrast, using the Cosmic Dawn (CoDa) III radiation hydrodynamical simulation, \citet{lewis_short_2022} found good agreement with the observed mean free path evolution.  Notably, the CoDa III simulation was performed at substantially better mass resolution (with a dark matter particle mass $M_{\rm dm}=5.09\times 10^{4}\,{\rm M}_{\odot}$ in a $94.4\, \rm cMpc$ box) compared to the \citet{keating_constraining_2020} model ($M_{\rm dm}=5.1\times 10^{7}\,{\rm M}_{\odot}$ in a $236\,\rm cMpc$ box).  However, CoDa III was not calibrated to reproduce the post-reionisation \Lya forest transmission.  Intermediate in scale between these two cases is the \textsc{thesan} radiation hydrodynamical simulation \citep{Garaldi_THESAN_2022}, with $M_{\rm dm}=3.12\times 10^{6}\rm\,M_{\odot}$ and a box size of $95.5\rm\,cMpc$. The redshift evolution of the mean free path in \textsc{thesan} was found to be in good agreement with the observational data, but with a normalisation that lies slightly above the $1\sigma$ uncertainty on the \citet{becker_mean_2021} and \citet{zhu_mfp_2023} measurements at $z=6$. In all three of these independent simulations, a late end to reionisation at $z<6$ was assumed.  Finally, the CROC radiation hydrodynamical simulations \citep{Gnedin2014} used a box size of $59\rm\,cMpc$ and a dark matter particle mass $M_{\rm dm}=7\times 10^{6}\,{\rm M}_{\odot}$.  \citet{Fan2024} recently demonstrated that CROC overshoots the mean free path measurements at $z\simeq 6$, most likely because the IGM is already highly ionised (with volume averaged \HI fraction $\langle x_{\rm  HI}\rangle < 10^{-3}$)  by $z\sim 6.4$ in these models.

One possible reason for differences between observation and theory is if the mean free path measurements at $z\simeq 6$ suffer a systematic bias.  The measurements presented by \citet{becker_mean_2021} and \citet{zhu_mfp_2023} at $z\simeq 6$ are determined from the Lyman continuum opacity in the vicinity of bright quasars.  Biases may arise if there is an over-correction for the quasar proximity effect \citep{daloisio_large_2018}, or an under-correction for the clustering of ionising photon sinks around the quasar host haloes \citep{Prochaska2014,Theuns_LLS_2024}.  However, \citet{Satyavolu_mfp_2023} and \citet{Roth_mfp_2023} have recently investigated a range of potential biases in the mean free path measurements, finding them to be well controlled within the analysis framework used by \citet{becker_mean_2021} and \citet{zhu_mfp_2023}. \citet{Satyavolu_mfp_2023} also found no significant difference in mean free path values when comparing the various definitions employed in the recent literature.

Another possibility is the simulations lack the dynamic range necessary for capturing the redshift evolution of the mean free path; the large simulation volumes needed to model the size distribution of ionised bubbles during reionisation \citep[e.g.][]{Iliev2014,Kaur2020} are too low resolution to simultaneously resolve small-scale photon sinks.  \citet{Cain2021} suggested that improved agreement with the mean free path measurements may be obtained by invoking a sub-grid model for the ionising photon sinks.  These sinks may be associated with unresolved Lyman-limit systems and/or mini-haloes with masses $<10^{8}\rm\,M_{\odot}$ \citep{Park2023}.  Furthermore, earlier work by \citet{Rahmati_mfp_2018} using the Aurora suite of radiation hydrodyanamical simulations\footnote{The fiducial Aurora simulation used a dark matter particle mass, $M_{\rm dm}=1\times 10^{7}\,{\rm M}_{\odot}$ and a box size $35.2\rm\,cMpc$.} highlighted the importance of \HI absorption systems with $N_{\rm HI}\sim 10^{16}$--$10^{17}\rm\,cm^{-2}$ for setting the mean free path during the final stages of reionisation (i.e., strong \Lya forest absorbers that arise from the diffuse, already reionised IGM, rather than dense neutral clumps).  \citet{Nasir_photon_sink_2021} similarly argued that gas that has dynamically responded to the increased thermal pressure from photo-heating will dominate the Lyman-limit opacity by $z\simeq 6$, and this gas will typically have \HI column densities $N_{\rm HI}<10^{17.2}\rm\,cm^{-2}$.  Hence, resolving structure in the diffuse IGM during the end stages of reionisation is a key requirement for capturing the evolution of the mean free path in simulations of inhomogeneous reionisation.

In this context, we examine the Lyman-limit photon mean free path predicted by a sub-set of the Sherwood-Relics\footnote{https://www.nottingham.ac.uk/astronomy/sherwood-relics/} simulations \citep{puchwein_sherwood-relics_2022}.  These are a large suite of simulations that model the effect of inhomogeneous reionisation   on the high-redshift IGM and \Lya forest, including the hydrodynamical response of intergalactic gas to patchy photoionisation and heating.  In contrast to other recent numerical simulation work, the Sherwood-Relics simulations use a hybrid approach that combines radiative transfer calculations performed using ATON \citep{aubert_radiative_2008} with P-Gadget-3 cosmological hydrodynamical simulations \citep{Springel_gadget_2005}.    In addition, because Sherwood-Relics was designed to study the small-scale structure in the \Lya forest \citep[e.g.][]{Molaro2023,Irsic2024}, the mass resolution of our fiducial model ($M_{\rm dm}=7.9\times 10^{5}\rm\,M_{\odot}$ in a $59\rm\,cMpc$ box, which lies between CoDa III and \textsc{thesan}) has been selected with this specific goal in mind \citep[cf.][]{BoltonBecker2009,Doughty2023}.

This paper is organised as follows.  In Section~\ref{modeling} we introduce the simulations used in this work and perform an initial comparison of the Lyman-limit photon mean free path predicted by Sherwood-Relics to the observational data.   We investigate the physical properties of the absorption systems that set the mean free path in Section~\ref{systems_that_set_mfp}, and examine the expected bias in the mean free path around the host haloes of ionising sources in Section~\ref{mfp_halos}.  Finally, we conclude in Section~\ref{conclusions}.  A set of appendices contain some numerical tests and a discussion of the analytical scaling relations commonly used for modelling the Lyman-limit opacity.

%%%%%%%%%%%%%%%%%%%%%%%%%%%%%%%%%%%%%%%%%%%%%%%%%%%%%%%%%%%%%%%%%%
%%%%%%%%%%%%%%%%%%%%%%%%%%% SECTION 2 %%%%%%%%%%%%%%%%%%%%%%%%%%%%
%%%%%%%%%%%%%%%%%%%%%%%%%%%%%%%%%%%%%%%%%%%%%%%%%%%%%%%%%%%%%%%%%%

\section{Modelling the mean free path with Sherwood-Relics} \label{modeling}
\subsection{Hydrodynamical simulations}

\begin{table*}
    \centering
    \begin{tabular}{c|c|c|c|c|c|c|c}
    \hline
      Name  &  $L_{\rm box}$ $[h^{-1}\rm\,cMpc]$ & N$_{\rm part}$ & $l_{\rm soft}$ $[h^{-1}\,\rm ckpc]$ & $M_{\rm dm}$  [$h^{-1}$M$_\odot$ ]&M$_{\rm gas}$  [$h^{-1}$M$_\odot$ ]& $z_{\rm r}$ & $z_{\rm mid}$ \\ \hline
      
      40-2048 & 40 & $2\times 2048^3$ & 0.78 & $5.37 \times 10^5$ & $9.97 \times 10^4$ &  5.7 & 7.5  \\ 
      160-2048 & 160 & $2\times 2048^3$ & 3.13 &  $3.44 \times 10^7$ & $6.38 \times 10^{6}$ & 5.3 & 7.2 \\ 
      40-1024 & 40 & $2 \times 1024^3$ & 1.56 & $4.30\times 10^6$ &  $7.97 \times 10^5$ & 5.7 & 7.5 \\
      40-512 & 40 & $2\times 512^3$ & 3.13 &  $3.44 \times 10^7$ & $6.38 \times 10^{6}$ & 5.7 & 7.5 \\ 
      \hline
    \end{tabular}
    \caption{Summary of the Sherwood-Relics simulations \citep{puchwein_sherwood-relics_2022} used in this work.  The columns, from left to right, list the model name, box size in $h^{-1}\,\rm cMpc$, the gas and dark matter particle number, the gravitational softening length in $h^{-1}\,\rm ckpc$, the dark matter and gas particle masses in $h^{-1}M_{\odot}$, the redshift, $z_{\rm r}$, where the volume averaged neutral hydrogen fraction first falls below $\langle x_{\rm HI}\rangle = 10^{-3}$, and the mid-point of reionisation, $z_{\rm mid}$, where $\langle x_{\rm HI}\rangle=0.5$.}
    \label{tab:models}
\end{table*}

\begin{figure*}
    \centering
\includegraphics[width=\textwidth]{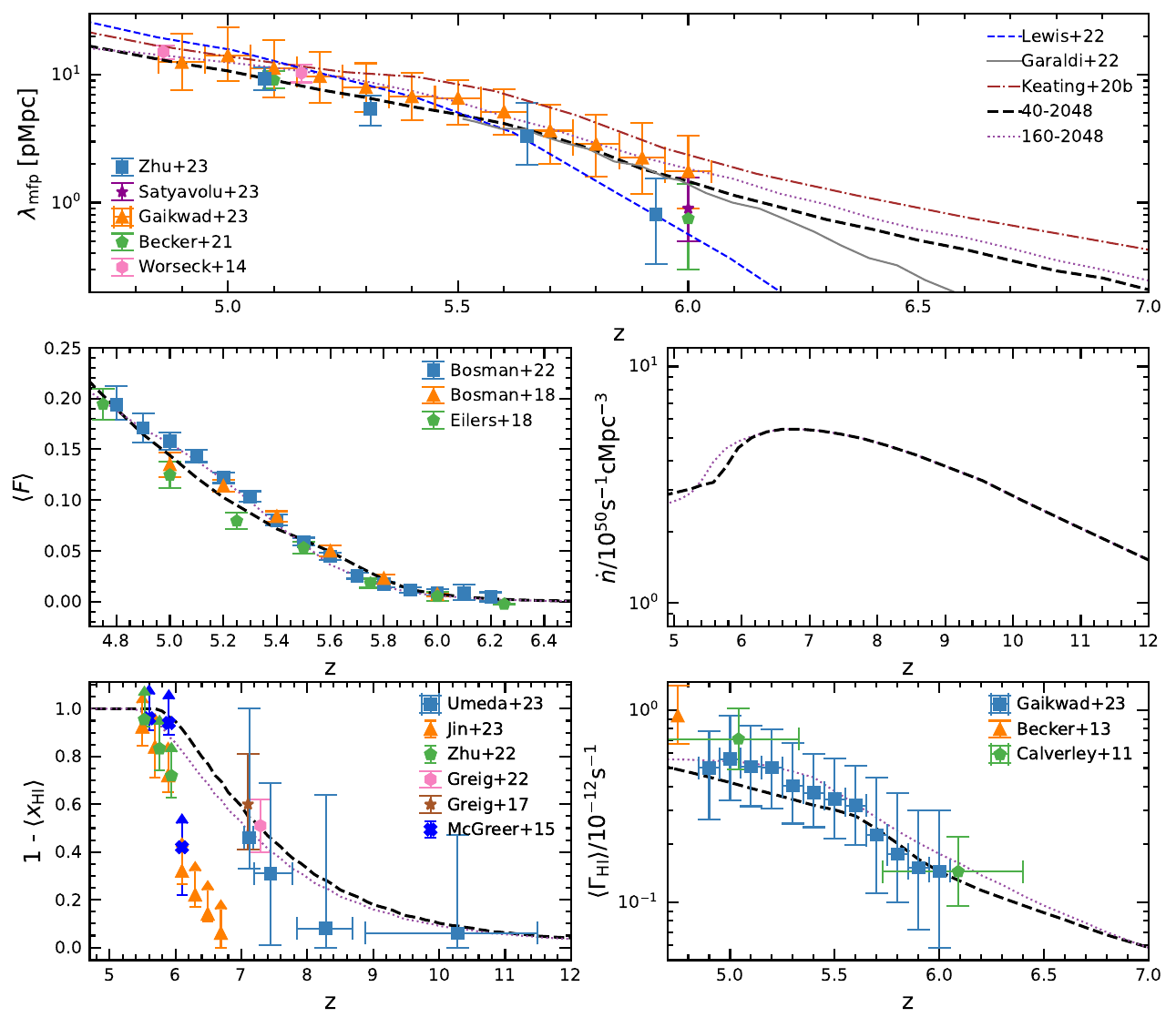}
    \vspace{-0.7cm}
    \caption{Comparison of two simulations used in this work, 40-2048 (black dashed curves) and 160-2048 (violet dotted curves) to observational data.  {\it Top:} The Lyman-limit photon mean free path.  The observational data are obtained from the Lyman-limit transmission in stacked quasar spectra \citep{Prochaska_opacity_2009,Worseck2014,becker_mean_2021,zhu_mfp_2023} and (indirectly) from the \Lya forest \citep{gaikwad_measuring_2023}. All error bars are $1\sigma$.  Predictions from other recent simulations are the low $\tau_{\rm CMB}$ model in \citet{keating_constraining_2020} (brown dot-dashed curve), \textsc{thesan-1} \citep[][solid grey curve]{Garaldi_THESAN_2022} and CoDa III \citep[][blue dashed curve]{lewis_short_2022}.  {\it Middle left:} The mean transmission of the Ly-$\alpha$ forest against redshift \protect\citep{eilers_opacity_2018,bosman_new_2018,bosman_hydrogen_2022}. The 40-2048 model was calibrated to match \protect\cite{eilers_opacity_2018} and \protect\cite{bosman_new_2018}, whereas the 160-2048 simulation has been calibrated to reproduce \protect\cite{bosman_hydrogen_2022}.  The good agreement with the data at $z>5$ is by design and is not a prediction. {\it Middle right: The ionising emissivity against redshift.} {\it Bottom left:}  The volume averaged ionised hydrogen fraction.   The observational data include lower limits from dark pixels and gaps in the Ly-$\alpha$ and Ly-$\beta$ forest \protect\citep{mcgreer_model-independent_2015,zhu_dark_gaps_2022,jin_nearly_2023} and the \Lya damping wing in quasar \protect\citep{greig_are_2016,greig_igm_2022} and galaxy \citep{Umeda_nhi_2023} spectra.  {\it Bottom right:} The volume averaged \HI photoionisation rate.  Observational constraints are from the quasar proximity effect \citep{calverley_measurements_2011} and the \Lya forest transmission \citep{becker_new_2013,gaikwad_measuring_2023}. \emph{Note the redshift range on the horizontal axis is not the same in all panels.}}
    \label{fig:overview_figure}
\end{figure*}

\begin{table}
    \centering
    \begin{tabular}{c|c|c|c|c}
    \hline
      $z$  &  $\langle x_{\rm HI}\rangle$ & $\lambda_{\rm mfp}$ [pMpc] &  $\langle x_{\rm HI}\rangle$ & $\lambda_{\rm mfp}$ [pMpc] \\ 
      & (40-2048) & (40-2048) & (160-2048) & (160-2048) \\
      \hline
      5.4 & $10^{-4.19}$ & 5.63 & $10^{-2.35}$ & 7.44 \\
      6.0 & 0.05 & 1.46  & 0.14 & 1.84 \\
      7.0 & 0.41 & 0.20 & 0.48 & 0.25 \\
      8.0 & 0.67 & 0.05 & 0.71 & 0.05 \\
      10.0 & 0.90 & $10^{-2.17}$ & 0.91 & $10^{-2.24}$ \\
      \hline
    \end{tabular}
    \caption{The volume averaged neutral hydrogen fraction, $\langle x_{\rm HI}\rangle$, and mean free path for Lyman-limit photons, $\lambda_{\rm mfp}$, predicted by the 40-2048 and 160-2048 simulations, at selected redshifts, $z$.}
    \label{tab:MFP}
\end{table}

In this work we use a sub-set of the Sherwood-Relics simulations, described in detail in \citet{puchwein_sherwood-relics_2022}.  These are a suite of high-resolution cosmological hydrodynamical simulations performed with a modified version of the P-Gadget-3 code \citep{Springel_gadget_2005}.  We use cosmological boxes with size $40h^{-1}\rm\,cMpc$ and $160 h^{-1}\rm\,cMpc$, each with $2 \times 2048^3$ dark matter and gas particles.  We refer to these simulations as 40-2048 and 160-2048, respectively (see Table~\ref{tab:models}).  The 160-2048 model uses the same mass resolution, box size and initial conditions as the (hydrodynamically decoupled) ATON simulations presented by \citet{Keating2020,keating_constraining_2020}, although in this work we assume a slightly different reionisation history. In addition, we use two smaller simulations of size $40h^{-1}\rm\,cMpc$ with $2 \times 512^3$ and $2 \times 1024^3$ dark matter and gas particles, which we refer to as 40-512 and 40-1024, respectively.  These models are identical to the 40-2048 run except  for particle masses that are $64$ and $8$ times larger.  The mass resolution of the 40-512 model matches the mass resolution of 160-2048.

The Sherwood-Relics simulations follow inhomogeneous reionisation using a novel hybrid approach.  We refer the reader to \citet{puchwein_sherwood-relics_2022} for further details on the numerical scheme.  In brief, the radiative transfer of monochromatic, ultra-violet photons in the simulations is followed using the moment-based M1-closure radiative transfer code ATON \citep{aubert_radiative_2008,Aubert2010}.  Rather than performing full radiation-hydrodynamical (RHD) simulations, however, we take an intermediate approach, where pre-generated three-dimensional maps of the \HI photoionisation rates from ATON are applied on-the-fly to the hydrodynamical simulations.  This has the advantage of self-consistently following the pressure response of the intergalactic gas to reionisation, but without the computational overhead of a full RHD calculation that also incorporates a detailed sub-grid model for star formation and feedback.

Gas particles with temperature $T<10^{5}\rm\,K$ and density $\Delta=\rho/\langle \rho \rangle >1000$ are immediately converted into star particles in all models \citep[the "quick Ly-$\alpha$" approximation,][]{Viel2004}.  The simulations will therefore not correctly capture the incidence of very high density gas in haloes, but this will only affect the mean free path once the Lyman-limit opacity is dominated by gas within the virial radius of haloes (i.e., well after reionisation has completed). A test of this "quick Ly-$\alpha$" approximation and its effect on the mean free path is provided in Appendix~\ref{app:qLya}.  The mass resolution of our main simulation, 40-2048, has been chosen to adequately resolve small-scale structure in the low density IGM, while also maintaining a large enough cosmological volume to sample a good range of halo masses.  We will use the 160-2048 and 40-512 models to assess the effect of simulation volume and mass resolution on our results.  For all simulations, a $\Lambda$CDM cosmology with $\Omega_{\rm m}=0.308$, $\Omega_\Lambda=0.692$, $\Omega_{\rm b}=0.0482$, $h=0.678$, $\sigma_8=0.829$ and $n=0.961$ is assumed \citep{planck2014}.

The luminosity of \HI photo-ionising sources in the simulations is proportional to the total halo mass, where a minimum host halo mass of $M_{\rm h} > 10^9 h^{-1} {\rm M}_\odot$ is assumed.  The mean energy for the ionising photons is $18.6\rm\,eV$, corresponding to a black body spectrum with $T=40,000\rm\,K$.  Following the approach used in earlier work by \citet{Kulkarni2019} and \citet{Keating2020}, two of the models we use here have been calibrated to match observational measurements of the mean transmitted flux of the \Lya forest.  The 40-2048 simulation was calibrated to match the results from \cite{bosman_new_2018} and \cite{eilers_opacity_2018}, whereas the 160-2048 simulation was calibrated to match the more recent \cite{bosman_hydrogen_2022} measurements.  Despite the overlap between the different \Lya forest measurements, however, the 40-2048 and 160-2048 simulations have slightly different reionisation histories, with mid-points (i.e., where the volume average neutral hydrogen fraction $\langle x_{\rm HI}\rangle=0.5$) at $z_{\rm mid}=7.5$ and $z_{\rm mid}=7.2$, respectively.   This difference is attributable to the different simulation volumes (see Section~\ref{mfp_calc}). The 40-512 and 40-1024 models use the same reionisation history as the 40-2048 model.

\subsection{Comparison of the simulated mean free path to observations}
\label{mfp_calc}

A variety of approaches have been used to compute the Lyman-limit mean free path in the literature \citep[see][for a recent discussion]{Satyavolu_mfp_2023}.  In this work we follow \citet{Rahmati_mfp_2018} and use the mean of the free path distribution obtained from the simulation volume. \citet{Satyavolu_mfp_2023} demonstrated this yields a mean free path in good agreement with the e-folding scale obtained by fitting the average Lyman continuum transmission with an exponential \citep[e.g.,][]{Prochaska_opacity_2009}.

The free path, $\lambda_{\rm fp}$, is defined as the distance over which a Lyman-limit photon ($\lambda=912\,\angstrom$) will encounter an optical depth, $\tau_{\rm 912}$, equal to unity.  For a sight-line, $i$, drawn a proper distance $R$ parallel to the Cartesian axes of the periodic simulation volume, we require
\begin{equation} 
\label{eq:tau_iter}
\tau_{\rm 912,i} = \sigma_{912} \int_{0}^{\lambda_{\rm fp,i}} n_{\rm HI}(R)\,dR =1,  
\end{equation}
where $\sigma_{912}=6.348\times 10^{-18}\rm\,cm^{2}$ is the photoionisation cross-section \citep{Verner_cross_1996}.  We choose the starting position of each sight-line in the simulation at random.  If the distance required to achieve an optical depth of unity exceeds the box size, we add further randomly selected sight-lines onto the first.  The mean free path, $\lambda_{\rm mfp}$, is then the average free-path over an ensemble of $N$ sight-lines
\begin{equation} 
\label{eq:mfp}
\lambda_{\rm mfp} \equiv \langle \lambda_{\rm fp,i}\rangle = \frac{1}{N}\sum_{\rm i=0}^{\rm N}\lambda_{\rm fp,i}.
\end{equation}
Throughout this work we use between $N=10^{4}-10^6$, which provides a well converged measurement.

The 40-2048 and 160-2048 simulations are compared to observational constraints in Fig.~\ref{fig:overview_figure}. Clockwise from the top panel, these are: the mean free path, $\lambda_{\rm mfp}$, for Lyman limit photons,  the ionising emissivity, $\Dot{n}$, the volume averaged \HI photoionisation rate, $\langle \Gamma_{\rm HI}\rangle$, the volume averaged ionised hydrogen fraction, $1-\langle x_{\rm HI}\rangle$, and the mean transmission in the \Lya forest, $\langle F \rangle$.  The good agreement with the observed mean \Lya forest transmission at $z>5$ in the middle left panel is by design, and is not a prediction of the models.   Both simulations have a late end to reionisation, at $z<6$ (see Table~\ref{tab:models}), and are consistent with observational measurements of the ionised hydrogen fraction and the \HI photoionisation rate at $z>5$. The ionising emissivity drops to $\Dot{n}\simeq 3\times 10^{50}\rm\,s^{-1}\rm\,cMpc^{-3}$ at $z<6$ in both models, where $\Dot{n}\propto \Gamma_{\rm HI}/\lambda_{\rm mfp}$.  This behaviour is qualitatively similar to other recent reionisation models  \citep{Kulkarni2019,Keating2020,keating_constraining_2020,Cain2021,gaikwad_measuring_2023,Asthana2024} and is driven by mandating the simulations reproduce the observed \Lya forest transmission.  An ionising emissivity of $\Dot{n}\simeq 3\times 10^{50}\rm\,s^{-1}\rm\,cMpc^{-3}$ corresponds to only $1$--$2$ ionising photons per hydrogen atom emitted over the age of the Universe at $z=6$ \citep{MiraldaEscude2003,BoltonHaehnelt2007}.  Given that the collapsed mass fraction in haloes will monotonically rise toward lower redshift, the apparent need for this rather low ionising emissivity at $z=6$ implies some redshift evolution in the escape fraction or efficiency of ionising photon sources at $z>6$ \citep{Ocvirk2021}. Alternatively, models of the IGM approaching the end stages of reionisation may still be incomplete \citep[see][for a recent discussion]{Cain_output_2023}.

\begin{figure}
    \includegraphics[width=0.48\textwidth]{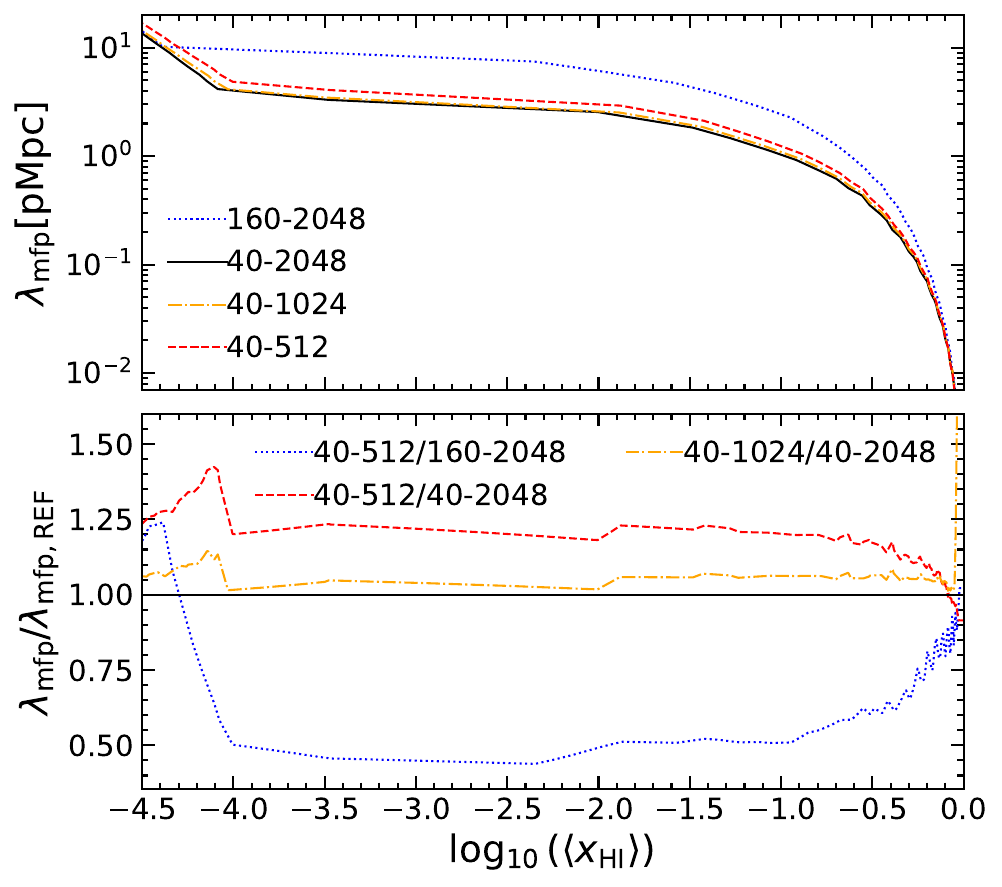}
    \vspace{-0.6cm}
    \caption{{\it Top:} The mean free path, $\lambda_{\rm mfp}$, against the volume averaged \HI fraction, $\langle x_{\rm HI}\rangle$, for the Sherwood-Relics simulations used in this work.  The 40-2048 (black solid curve), 40-1024 (orange dash dotted curve) and 40-512 (red dashed curve) models have the same box size, $40h^{-1}\rm\,cMpc$, but the gas particle mass in 40-512 (40-1024) is $64$ ($8$) times larger than 40-2048, which has $M_{\rm gas}=9.97\times 10^{4}h^{-1}\,\rm M_{\odot}$ (see Table~\ref{tab:models}).  The 160-2048 simulation (blue dotted curve, box size $160h^{-1}\rm\,cMpc$) has $64$ times the volume of 40-2048 but has the same mass resolution as 40-512. {\it Bottom:} The ratio of the mean free path at fixed box size for a factor of $64$ and $8$ difference in the gas particle mass (red dashed and orange dash dotted curves, respectively), and at fixed mass resolution for a factor of $64$ difference in simulation volume (blue dotted curve). The sharp features at $\log_{10}\langle x_{\rm HI}\rangle \sim -4$ arise from differences in the neutral fraction where the IGM transitions to being fully reionised.}
    \label{fig:mass_res}
\end{figure}

The redshift evolution of the mean free path in the simulations (see also Table~\ref{tab:MFP}) is in good agreement (within $1\sigma$) with the recent (indirect) determination of the mean free path from the \Lya forest transmission at $4.9<z<6$ \citep{gaikwad_measuring_2023}.   Similar $\lambda_{\rm mfp}$ measurements have also recently been independently reported by \citet{davies_constraints_2023}.  The models are also in good agreement with the mean free path measured from the Lyman continuum transmission in stacked quasar spectra at $z<5.7$ \citep{Worseck2014,becker_mean_2021,zhu_mfp_2023}.  For the highest redshift data point at $z=5.93$, the 40-2048 (160-2048) model is $1.2\sigma$ ($1.8\sigma$) above the recent \citet{zhu_mfp_2023} measurement.  At $z<6$, the mean free path redshift evolution for the 40-2048 model is very similar to \textsc{thesan-1} \citep[][solid grey curve]{Garaldi_THESAN_2022}, but more gradual than CoDa III \citep[][blue dashed curve]{lewis_short_2022}.  Note also the mean free path from \citet{keating_constraining_2020} is larger than the 160-2048 simulation at $z>5.2$, despite these models both using P-Gadget-3 and ATON simulations performed at the same mass resolution and box size.  This is in part because \citet{keating_constraining_2020} adopted a larger emissivity at $z<7$ by calibrating to match the \citet{bosman_new_2018} \Lya forest transmission.  However, another difference is that \citet{keating_constraining_2020} used ATON simulations that did not follow the patchy hydrodynamical response of the gas to photo-heating during reionisation.  Instead, they performed post-processed radiative transfer simulations on IGM density fields extracted from a cosmological simulation that was quickly reionised and heated at $z=15$ by the spatially uniform \citet{HaardtMadau2012} UV background.  This early, uniform heating of the IGM will assist in suppressing small-scale gas clumping and could also increase the mean free path.

Furthermore, at $5\leq z \leq 7$ -- where the mean \Lya forest transmission in the two Sherwood-Relics simulations in Fig.~\ref{fig:overview_figure} is similar by design -- the mean free path in the 160-2048 model is $\sim 20$ per cent larger than 40-2048.  This is despite reionisation being slightly delayed (by $\Delta z \sim 0.3$--$0.4$) in 160-2048 relative to the 40-2048 simulation.  This difference is partly because the 160-2048 model contains more massive haloes and larger, rarer ionised bubbles, as a result of the $64$ times larger simulation volume \citep[cf.][]{Iliev2014,Kaur2020,Lu2024}.  However, another difference is mass resolution; the 160-2048 simulation has a factor of $64$ poorer mass resolution compared to 40-2048 (see Table~\ref{tab:models}).  The 160-2048 simulation under-resolves the Lyman-limit opacity from small scale structure in the \emph{already reionised} IGM.

We examine the effect of simulation mass resolution and box size on the mean free path further in Fig.~\ref{fig:mass_res}. Here we choose to show the mean free path against the volume averaged \HI fraction, $\langle x_{\rm HI} \rangle$, rather than against redshift.  This allows us to isolate the effect of box size and mass resolution from the redshift dependent ionisation state of the IGM \citep[see e.g.][]{Rahmati_mfp_2018}.  The mean free path in the 40-512 (40-1024) simulation is typically 20 (5) per cent larger than the 40-2048 model, and is as much as 40 (10) per cent larger at $\langle x_{\rm HI} \rangle=10^{-4}$.  This is due to unresolved small-scale structure in the lower resolution models, resulting in an underprediction of the Lyman-limit opacity from the reionised IGM. The mean free path in the 40-2048 simulation should therefore be reasonably well converged (within $5$ per cent) with mass resolution.  However, the mean free path in the 160-2048 simulation will not be converged with mass resolution and is likely overestimated by \emph{at least} 20 per cent. Hence, as has been discussed elsewhere \citep[e.g.][]{Cain2021,Satyavolu_mfp_2023,Park2023,Georgiev2024}, part of the reason for previous models failing to fully capture the redshift evolution of the mean free path first reported by \citet{becker_mean_2021} may be unresolved photon sinks.  We argue in Section~\ref{sec:opacity} these sinks include residual neutral hydrogen in the diffuse, reionised IGM. 

In contrast, the smaller box size of $40h^{-1}\rm\,cMpc$ used for the 40-512 simulation results in an underprediction of the mean free path by up to a factor of two, relative to 160-2048.  This is because of the larger ionised bubbles and brighter ionising sources / more massive haloes present in the 160-2048 simulation.  This explains why a slightly later end to reionisation (by $\Delta z = 0.3$--$0.4$ when compared to 40-2048) is required for the 160-2048 simulation to match the \Lya forest transmission. For comparison, \citet{Iliev2014} reported a box size of $>100h^{-1}\rm\,cMpc$ is required to converge on the mean reionisation history, and a box size of $>200h^{-1}\rm\,cMpc$ is needed to adequately capture the large-scale patchiness of reionisation.  Our 160-2048 model should therefore be marginally converged with box size.

%%%%%%%%%%%%%%%%%%%%%%%%%%%%%%%%%%%%%%%%%%%%%%%%%%%%%%%%%%%%%%%%%%
%%%%%%%%%%%%%%%%%%%%%%%%%%% SECTION 3 %%%%%%%%%%%%%%%%%%%%%%%%%%%%
%%%%%%%%%%%%%%%%%%%%%%%%%%%%%%%%%%%%%%%%%%%%%%%%%%%%%%%%%%%%%%%%%%

\section{The Lyman-limit opacity in Sherwood-Relics} \label{systems_that_set_mfp}
\subsection{Physical origin of the free-path distribution}

\begin{figure}
    \includegraphics[width=0.48\textwidth]{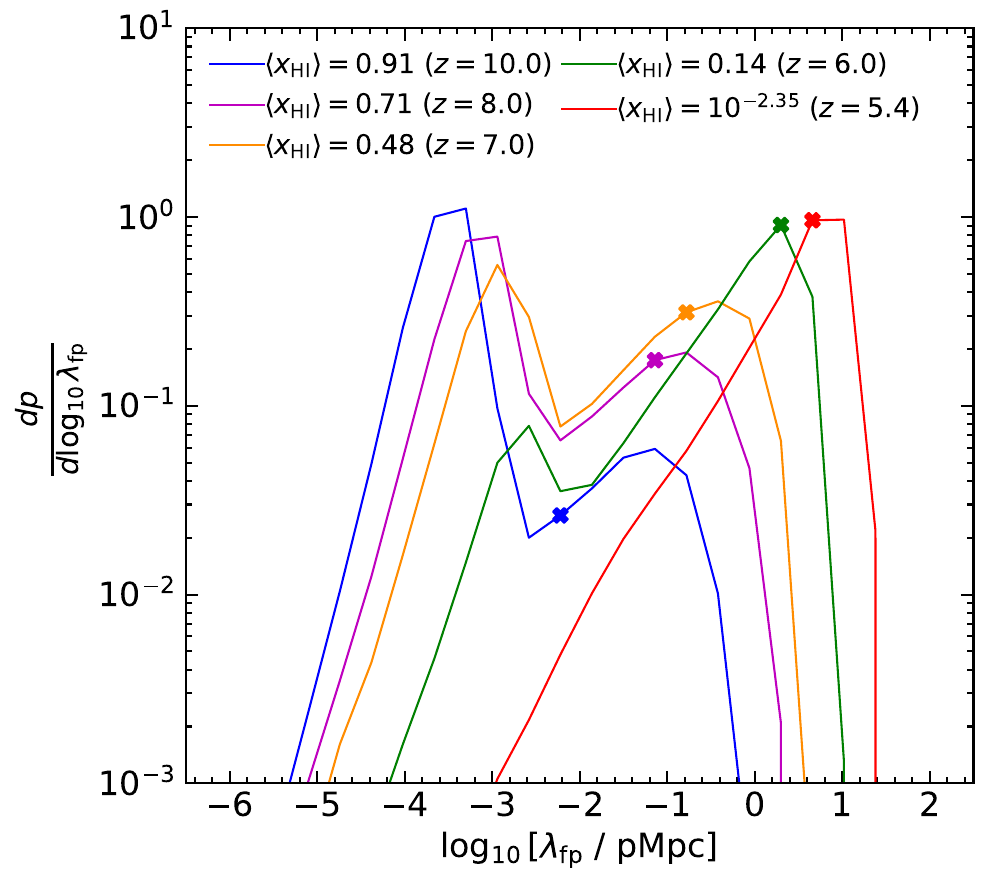}
    \vspace{-0.6cm}
    \caption{The probability density function of the logarithm of the Lyman-limit free path, $\log_{10}(\lambda_{\rm fp}/\rm pMpc)$, in the 40-2048 model for five different volume weighted neutral hydrogen fractions/redshifts (see Table~\ref{tab:MFP}).  The mean free path is shown as a cross on each distribution.  The bimodality arises from the distribution of ionised and neutral regions in the model and depends on the average \HI fraction. }
    \label{fig:pdf_over_a_range_of_z}
\end{figure}

\begin{figure*}
\centering
  \centering
    \includegraphics[width=0.85\textwidth]{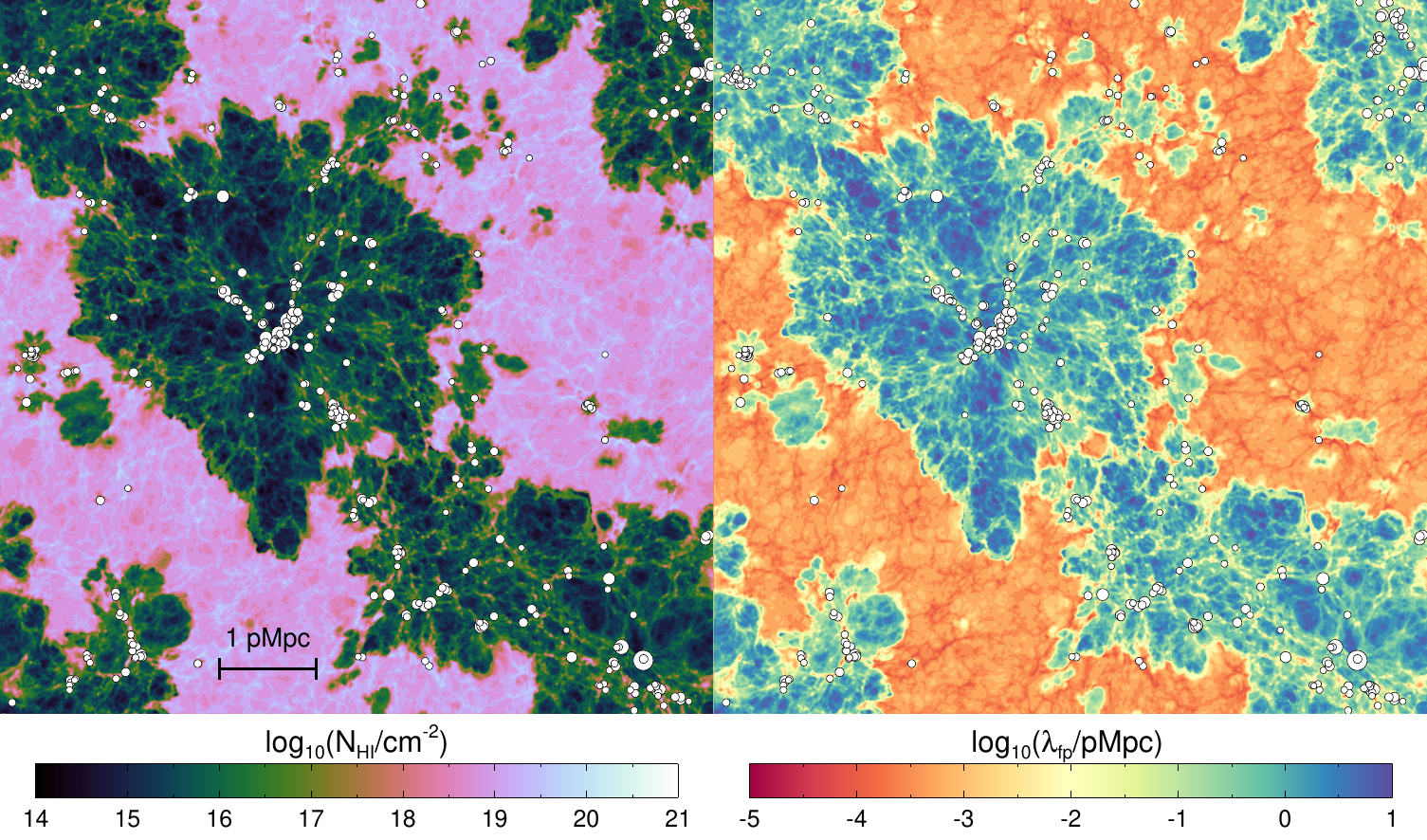}
    \vspace{-0.3cm}
    \caption{Projection of the logarithm of \HI column density, $\log_{\rm 10}(N_{\rm HI}/\rm cm^{-2})$ (left panel), and logarithm of the proper free path for Lyman-limit photons, $\log_{\rm 10}(\lambda_{\rm fp}/\rm pMpc)$ (right panel), in the 40-2048 model at $z=7$, when $\langle x_{\rm HI}\rangle=0.41$. Note that for this visualisation the free path has been estimated using the local \HI density in each pixel, rather than using Eq.~(\ref{eq:tau_iter}). The slice width is $64\rm\,pkpc$ and is $40h^{-1}\rm\,cMpc$ on a side.  The locations of ionising source host haloes (i.e., all haloes with total mass $M_{\rm h}\geq 10^{9}h^{-1}\rm\,M_{\odot}$) are shown as white circles, with sizes that scale with the logarithm of the halo mass.  The most massive halo in the slice -- located in the large ionised region slightly left of the box centre -- has total mass $M_{\rm h}=10^{11.25}h^{-1}\rm\,M_{\odot}$.}
    \label{fig:mfp_slice}
\end{figure*}

\begin{figure}
    \centering
    \includegraphics[width=0.45\textwidth]{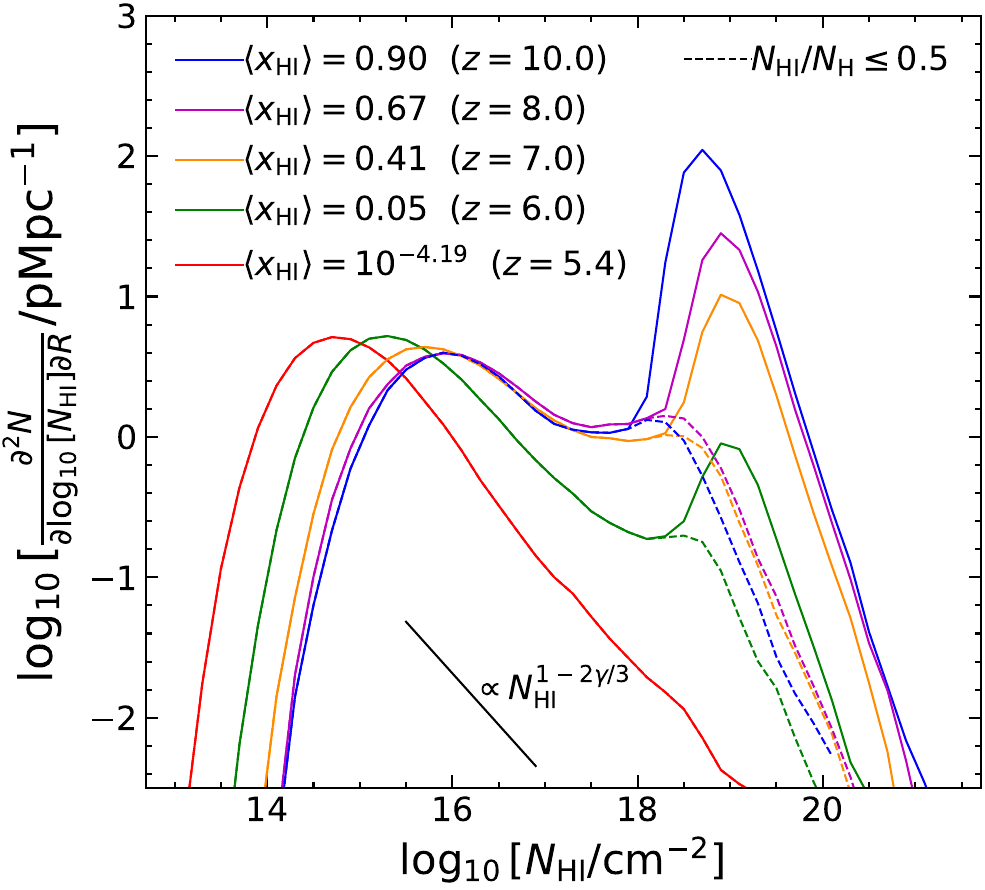} 
    \vspace{-0.2cm}
    \caption{The solid curves show the \HI column density distribution for the five different volume weighted \HI fractions/redshifts used in Fig.~\ref{fig:pdf_over_a_range_of_z} and summarised in Table~\ref{tab:MFP}. The dashed curves exclude column densities with $N_{\rm HI}/N_{\rm H}>0.5$, and therefore ignore gas that is yet to be reionised.  The diagonal black line, $f(\log_{10}N_{\rm HI},R) \propto N_{\rm HI}^{1-2\gamma/3}$, gives the expected slope of the post-reionisation \HI column density distribution assuming an underlying gas density distribution with a power-law high density tail, $P(\Delta)\propto \Delta^{-\gamma}$, with $\gamma=2.6$ (see text and Appendix~\ref{sec:scaling} for details).}
    \label{fig:column_density_threhold}
\end{figure}

\begin{figure*}
\centering
  \centering
  \includegraphics[width=1.0\textwidth]{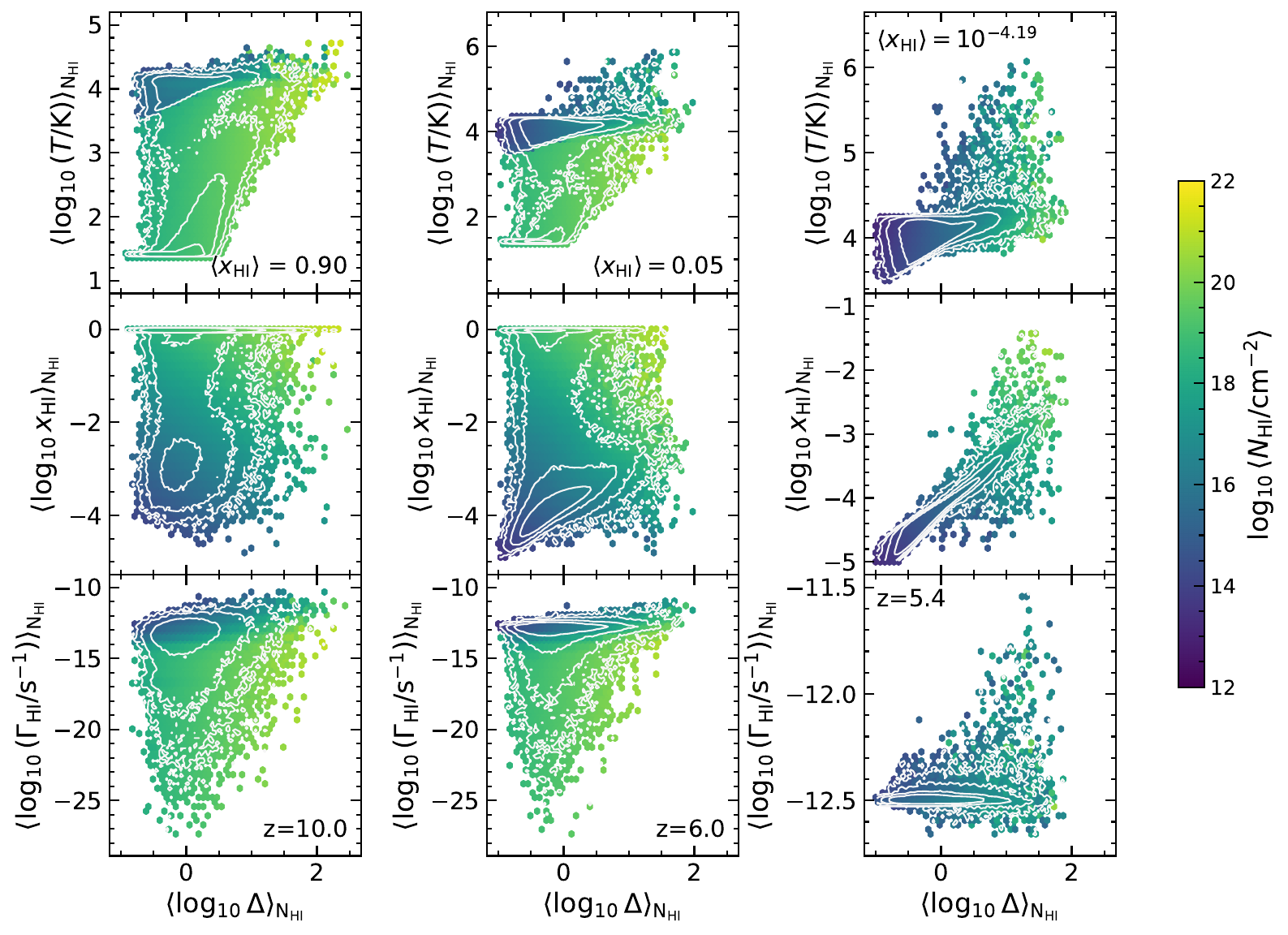}
  \vspace{-0.7cm}
  \caption{Each column in this figure shows a different volume averaged neutral fraction/redshift, where from left to right, $\langle x_{\rm HI}\rangle=0.90,\,0.05$ and $10^{-4.2}$, corresponding to $z=10,\,6$ and $5.4$ (see Table~\ref{tab:MFP}). {\it Top row:} A 2D histogram of the logarithm of the average temperature, $\langle \log_{10}(T/\rm K)\rangle_{\rm N_{\rm HI}}$, against the logarithm of the average overdensity, $\langle \log_{10}\Delta\rangle_{\rm N_{\rm HI}}$, within the integration windows used to calculate the \HI column densities in the 40-2048 simulation.    The value of the average \HI column density in each bin is displayed in the colour bar on the right of the figure.  Contours show the number density of systems in each bin, with the density increasing by $1$ dex for each contour level.  {\it Middle row:} As for the top row, but now showing the logarithm of the average \HI fraction in each integration window, $\langle \log_{10} x_{\rm HI}\rangle_{\rm N_{\rm HI}}$ against $\langle \log_{10}\Delta\rangle_{\rm N_{\rm HI}}$.  {\it Bottom row:} As for the top row, but now showing the logarithm of the average \HI photoionisation rate in each integration window, $\langle \log_{10} (\Gamma_{\rm HI}/\rm s^{-1}) \rangle_{\rm N_{\rm HI}}$ against $\langle \log_{10}\Delta\rangle_{\rm N_{\rm HI}}$. \emph{Note the different range on the vertical axis of each row.}} 
  \label{fig:hexplots}
\end{figure*}%

\begin{figure*}
\centering
  \centering
  \includegraphics[width=\textwidth]{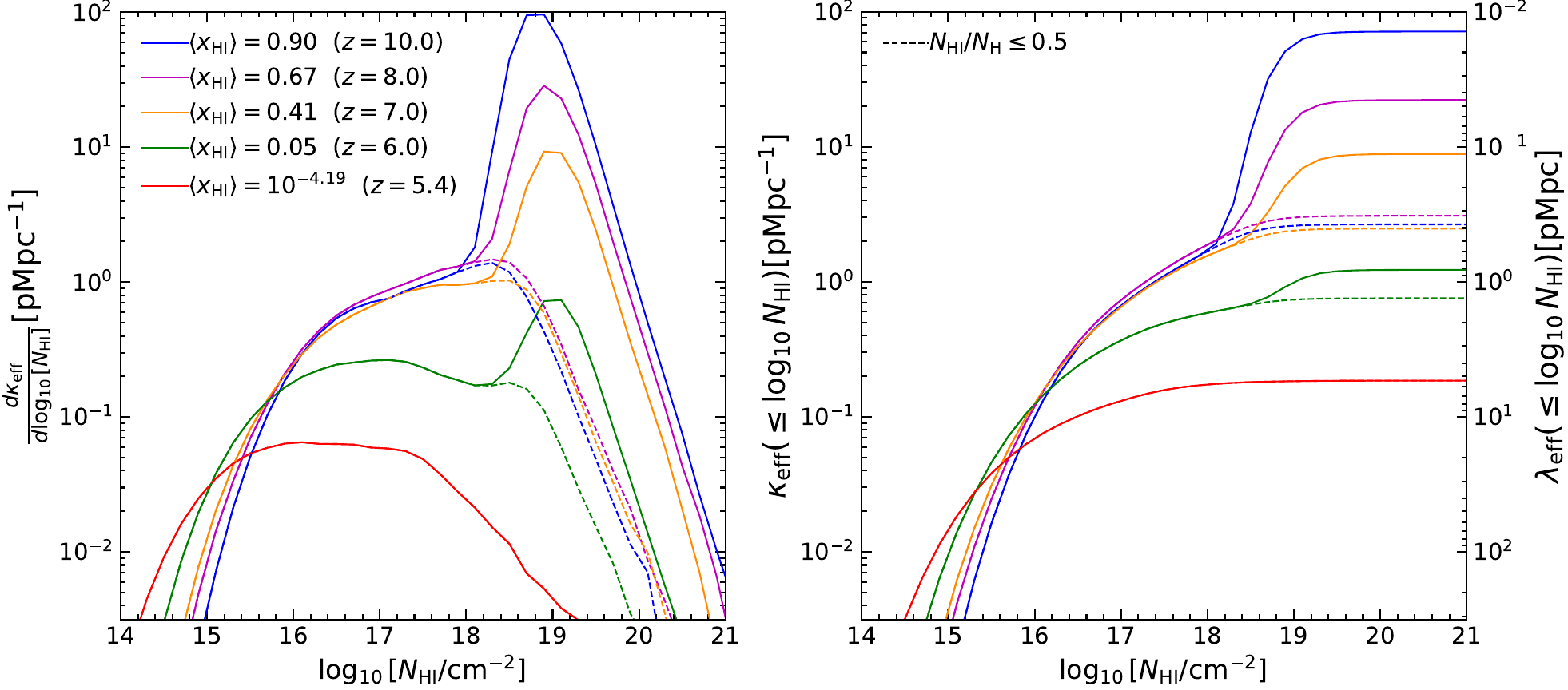}
  \vspace{-0.5cm}
  \caption{{\it Left:} The differential distribution of the Lyman-limit opacity, $\kappa_{\rm eff}$, as a function of $\log_{10}(N_{\rm HI}/\rm cm^{-2})$ from the 40-2048 model. As for Fig.~\ref{fig:pdf_over_a_range_of_z} and Fig.~\ref{fig:column_density_threhold}, five different volume weighted \HI fractions/redshifts are displayed (see Table~\ref{tab:MFP}).  The dashed curves are computed ignoring all \HI column densities with $N_{\rm HI}/N_{\rm H}>0.5$. {\it Right:} The corresponding cumulative distribution of the Lyman-limit opacity.  At $\langle x_{\rm HI} \rangle = 0.05$ ($z=6$, green solid and dashed curves), $61$ per cent of the total Lyman-limit opacity arises from hydrogen that is already ionised (i.e., with $N_{\rm HI}/N_{\rm H}\leq 0.5$).} 
  \label{fig:opacity}
\end{figure*}

We now turn to examine the origin of the free path distribution in the 40-2048 simulation.  The free path probability density functions (PDFs) for five different volume averaged neutral hydrogen fractions, $\langle x_{\rm HI} \rangle = 0.90,\, 0.67,\, 0.41,\, 0.05,\, 10^{-4.19}$  (corresponding to redshifts $z=10,\,8,\,7,\,6,\,5.4$, see Table~\ref{tab:MFP}) are shown in Fig.~\ref{fig:pdf_over_a_range_of_z}.  The mean free path, $\lambda_{\rm mfp}$, is shown by the cross on each curve.

Prior to the completion of reionisation (i.e., for $\langle x_{\rm HI}\rangle \gtrsim 10^{-3}$), the free path distribution is bimodal, with a characteristic minimum between the two modes at $\lambda_{\rm fp}\sim 10^{-2}\rm\,pMpc$.  The mode at smaller (larger) $\lambda_{\rm fp}$ arises from the neutral (ionised) hydrogen in the simulation volume.  The relative amplitudes of the modes depend on the average neutral fraction, and are similar for $\langle x_{\rm HI} \rangle\sim 0.5$.  In all cases, the range of free path values varies over $4$--$5$ orders of magnitude.  This range reflects the distribution of gas densities in the simulation, with the smallest free paths often associated with the highest density neutral gas.   Following the completion of reionisation ($\langle x_{\rm HI}\rangle \lesssim 10^{-3}$), there are no neutral islands and the bimodality is absent.  These results are very similar to earlier independent work by \cite{lewis_short_2022} using the CoDa III radiation hydrodynamical simulation, suggesting a bimodal distribution for $\lambda_{\rm fp}$ is a generic prediction of inhomogeneous reionisation models.

The spatial distribution of the free path is illustrated in Fig.~\ref{fig:mfp_slice}.  This shows a projection of the 40-2048 model at $z=7$ ($\langle x_{\rm HI}\rangle=0.41$) over a slice of width $64\rm\,pkpc$ -- approximately the Jeans length at the mean background density (see Eq.~(\ref{eq:jeans}) in Appendix~\ref{sec:window_size}).  The left panel of Fig.~\ref{fig:mfp_slice} shows the logarithm of the \HI column density, $\log_{10}(N_{\rm HI}/\rm cm^{-2})$, while the right panel displays the logarithm of the free path, $\log_{10}(\lambda_{\rm fp}/\rm\,pMpc)$. Pink ($N_{\rm HI} \gtrapprox 10^{18}\rm\,cm^{-2}$) and orange $(\lambda_{\rm fp}\lessapprox 10^{-3}\rm\,pMpc$) shading corresponds to neutral hydrogen that has yet to be reionised.  The origin of the bimodal distribution in Fig.~\ref{fig:pdf_over_a_range_of_z} is now evident, with the ionised bubbles around clustered sources (shown as white circles) producing the free path mode at $\lambda_{\rm fp}\sim 1\rm\,pMpc$.  Note also that $N_{\rm HI}$ and $\lambda_{\rm fp}$ vary \emph{within} the ionised and neutral regions, due to the underlying gas density fluctuations in the cosmic web.

We calculate \HI column densities, $N_{\rm HI}$, by integrating the neutral hydrogen density over windows equal to the Jeans length at the mean background density.  The choice of integration window is somewhat arbitrary, but is motivated by the results of \citet{schaye2001} and \citet{Rahmati_mfp_2018}, who found the local Jeans length is close to the typical size of \HI absorbers following reionisation.    We also tried integrating over fixed windows of width $50\rm\,km\,s^{-1}$ \citep[e.g.][]{Gurvich2017}, but found this did not qualitatively change our results (see Appendix~\ref{sec:window_size} for more details). 

The resulting \HI column density distribution function, $f(\log_{10}N_{\rm HI},R)$, is displayed in Fig.~\ref{fig:column_density_threhold}.  This is defined as the number, ${\mathscr N}$, of integration windows with column density $\log_{10}N_{\rm HI}$ per proper distance $R$, such that
\begin{equation} 
\label{eq:cddf}
f(\log_{10}N_{\rm HI},R) = \frac{\partial^{2}{\mathscr N}}{\partial\log_{10}N_{\rm HI}\,\partial R}.
\end{equation}

\noindent
Similar to the free path distribution, the column density distribution in Fig.~\ref{fig:column_density_threhold} is bimodal if $\langle x_{\rm HI}\rangle > 10^{-3}$, with modes at $N_{\rm HI}\sim 10^{15.5}\rm\,cm^{-2}$ ($\sim 10^{19}\rm\,cm^{-2}$) arising from the predominantly ionised (neutral) hydrogen gas.   This is further highlighted by the dashed curves in Fig.~\ref{fig:column_density_threhold}, where all \HI column densities with $N_{\rm HI}/N_{\rm H}>0.5$ (i.e. hydrogen that is more than $50$ per cent neutral) have been removed from the distribution.  For the largest \HI fractions, the dashed curve deviates substantially from the solid curve at $N_{\rm HI}>10^{18}\rm\,cm^{-2}$.  Following the completion of reionisation, however, the dashed and solid curves are almost identical.  At column densities $N_{\rm HI}>10^{14}$--$10^{15}\rm\,cm^{-2}$ (i.e., where the distribution is complete), the shape of the distribution approximately follows a power-law for the predominantly ionised gas, although there is a hint of flattening in the slope at $N_{\rm HI}\sim 10^{17}\rm\,cm^{-2}$ due to self-shielded gas \citep[e.g.][]{Altay2011}.
When the \HI in the post-reionisation IGM is in photoionisation equilibrium, the shape of $f(\log_{10}N_{\rm HI},R)$ is directly related to the gas density distribution (see Appendix~\ref{sec:scaling}).  For a gas density distribution with a power-law tail, $P(\Delta)\propto \Delta^{-\gamma}$, we expect that $f(\log_{10}N_{\rm HI},R)\propto N_{\rm HI}^{1-2\gamma/3}$ \citep[e.g.][]{furlanetto2005, mcquinn2011}.

Fig.~\ref{fig:hexplots} shows 2D histograms that elucidate the relationship between the \HI column density and the gas density, temperature, \HI photoionisation rate and \HI fraction for $\langle x_{\rm HI} \rangle = 0.90,\, 0.41,$ and $10^{-4.19}$ (corresponding to $z=10,\,6,$ and $5.4$, respectively, in the 40-2048 simulation). Here we denote quantities that have been averaged over the column integration window with $\langle\ldots\rangle_{\rm N_{\rm HI}}$.  Prior to the completion of reionisation at $z<6$ the temperature and \HI fractions are bimodal, although note column densities with $\langle \Gamma_{\rm HI} \rangle_{\rm N_{\rm HI}}=0$ (i.e., for $\langle x_{\rm HI}\rangle_{\rm N_{\rm HI}}=1$) are not shown on the logarithmic scale used in the lower panels.  Column densities $N_{\rm HI}>10^{18}\rm\,cm^{-2}$ most often correspond to cool, $\langle T \rangle_{\rm N_{\rm HI}}\lesssim 10^{3}\rm\,K$ neutral gas, with photoionisation rates $\langle \Gamma_{\rm HI} \rangle_{\rm N_{\rm HI}}\lesssim 10^{-14}\rm\,s^{-1}$.  The largest column densities, $N_{\rm HI}>10^{21}\rm\,cm^{-2}$, are almost exclusively associated with the highest density gas with $\langle \Delta \rangle_{\rm N_{\rm HI}}\gtrsim 10$.  Gas that has been reionised has $\langle T \rangle_{\rm N_{\rm HI}}\sim 10^{4}\rm\,K$, $\langle \Gamma\rangle_{\rm N_{\rm HI}}\sim 10^{-12}$--$10^{-13}\rm\,s^{-1}$ and $\langle x_{\rm HI}\rangle_{\rm N_{\rm HI}}\lesssim 10^{-3}$, with typical column densities of $N_{\rm HI}\sim 10^{15}$--$10^{16}\rm\,cm^{-2}$ at the mean background density.  The upward scatter at $\langle T \rangle_{\rm N_{\rm HI}} > 10^{4.5}\rm\,K$, which is particularly evident in the left column of Fig.~\ref{fig:hexplots}, is associated with gas shocked by gravitational infall (see also fig. 5 of \citet{puchwein_sherwood-relics_2022}).

\subsection{Lyman-limit opacity as a function of \HI column density} \label{sec:opacity}

We now quantify the role that different \HI column densities play in determining the opacity of the IGM to Lyman-limit photons.  Here we follow \citet{Rahmati_mfp_2018} and \citet{Nasir_photon_sink_2021} and compute the Lyman-limit opacity by integrating over the \HI column density distribution \citep[see also][]{Prochaska_opacity_2009,becker_mean_2021,zhu_mfp_2023}.  The differential distribution of the opacity to Lyman-limit photons, $\kappa_{\rm eff}$, is calculated as
\begin{align} 
\frac{d\kappa_{\rm eff}}{d\log_{10}N_{\rm HI}} &= \frac{1}{\delta\log_{10}N_{\rm HI}} \nonumber \\
&\times  \int_{N_{\rm HI}}^{N_{\rm HI} + \delta N_{\rm HI}}  f(N_{\rm HI},R)(1 - e^{-N_{\rm HI}\sigma_{912}})\,dN_{\rm HI},
\end{align}
where $f(N_{\rm HI},R) = f(\log_{10} N_{\rm HI},R)/(N_{\rm HI} \ln 10)$ is the column density distribution function (i.e. Eq.~(\ref{eq:cddf})) and $\delta\log_{10}N_{\rm HI} = \log_{10}(1 + \delta N_{\rm HI}/N_{\rm HI})=0.2$ is our adopted bin size. The cumulative distribution of the opacity is then
\begin{equation}
    \kappa_{\rm eff}(\leq \log_{10} N_{\rm HI}) = \int_{0}^{\log_{10}N_{\rm HI}} \frac{d\kappa_{\rm eff}}{d \log_{10}N_{\rm HI}}\,d \log_{10}N_{\rm HI}.
\end{equation}
This quantity is related to the effective optical depth to Lyman-limit photons, $\tau_{\rm eff}$, by $\kappa_{\rm eff}=d\tau_{\rm eff}/dR$, and the attenuation length $\lambda_{\rm eff}$ by $\kappa_{\rm eff}=\lambda_{\rm eff}^{-1}$.  We find the attenuation length is typically within a factor of two of the mean free path given by Eq.~(\ref{eq:mfp}) \citep[see e.g. appendix A in][]{Theuns_LLS_2024}, but this factor will be sensitive to the choice of column density integration window.

The differential opacity is shown in the left panel of Fig.~\ref{fig:opacity}.  For $\langle x_{\rm HI}\rangle > 10^{-3}$ the differential opacity exhibits a peak at $N_{\rm HI}=10^{18}$--$10^{19}\rm\,cm^{-2}$ associated with mostly neutral hydrogen\footnote{Note that while column densities of $N_{\rm HI}=10^{18}$--$10^{19}\rm\,cm^{-2}$ are typically associated with Lyman-limit systems, prior to the end of reionisation these arise from \HI in the diffuse IGM, rather than from high density, self-shielded gas (see Fig.~\ref{fig:hexplots}).} with $N_{\rm HI}/N_{\rm H}>0.5$.  The peak occurs around $1$ dex higher compared to \citet{Rahmati_mfp_2018}, but as discussed earlier, this value will depend on our choice of integration window as well as differences in the assumed reionisation history, which for 40-2048 ends later than the fiducial model in \citet{Rahmati_mfp_2018}.  For $\langle x_{\rm HI}\rangle < 10^{-3}$ (i.e. at the end of reionisation), the contribution of column densities with $N_{\rm HI}>10^{18}$--$10^{19}\rm\,cm^{-2}$ to the total Lyman-limit opacity is minimal, and instead gas with $N_{\rm HI}\sim 10^{16}$--$10^{17}\rm\,cm^{-2}$ dominates the differential opacity (i.e. at the \HI column densities typical of strong \Lya forest absorbers). At $z=5.4$, these absorbers have average densities $\langle \Delta \rangle_{\rm N_{\rm HI}}\sim 3$, temperatures $\langle T\rangle_{\rm N_{\rm HI}}\sim 10^{4.1}\rm\,K$ and \HI fractions $\langle x_{\rm HI}\rangle_{\rm N_{\rm HI}}\sim 10^{-3.5}$.

Insight into the role the \emph{already reionised} IGM plays in setting the mean free path may be obtained from the cumulative opacity, displayed in the right panel of Fig.~\ref{fig:opacity}.   Here we report the fraction of the total opacity that originates from ionised regions with $N_{\rm HI}/N_{\rm H}\leq 0.5$ (the dashed curves in Fig.~\ref{fig:opacity}).  For the five neutral hydrogen fractions shown at $z=10,\,8,\,7,\,6$ and $5.4$, this represents $3,\,13,\,27,\,61$ and $>99$ per cent of the total opacity, respectively.  Hence, during the final stages of late reionisation at $z=6$,  \emph{the majority of the Lyman-limit opacity in the IGM arises from gas that is mostly ionised.}  This result is consistent with \citet{Rahmati_mfp_2018}, who pointed out the importance of the cumulative opacity arising from systems with $N_{\rm HI}<10^{16.5}\rm\,cm^{-2}$ (i.e., absorbers below the Lyman-limit system threshold of $N_{\rm HI}=10^{17.2}\rm\,cm^{-2}$) for setting the mean free path during the final stages of reionisation. These systems will be associated with reionised gas that may also have had time to dynamically respond to photo-heating \citep{Nasir_photon_sink_2021,puchwein_sherwood-relics_2022}.  Resolving the small-scale structure of the IGM at moderate overdensity is therefore a necessary requirement for modelling the mean free path during the final stages of reionisation. 

\begin{table}
    \centering
    \begin{tabular}{c|c|c|c|c}
    \hline
      $z$  & $\langle x_{\rm HI}\rangle $ &  $\langle M_{\rm h} \rangle$ $[h^{-1}\rm\,M_{\odot}]$ & $\Lambda_{\rm mfp}$ $[\rm pMpc]$ &$\Lambda_{\rm mfp}/\lambda_{\rm mfp}$  \\ 
      \hline
      5.4 & $10^{-4.19}$ & $10^{10.9}$ & 4.50 &  0.81 \\ 
     5.4 & $10^{-4.19}$ & $10^{10.1}$ & 4.59 &  0.83  \\
     5.4 & $10^{-4.19}$ & $10^{9.3}$ & 4.68 & 0.85 \\ 
     \hline
     6.0 & 0.05 & $10^{10.7}$ & 1.42 &  0.98 \\ 
      6.0 & 0.05 & $10^{10.0}$ & 1.30 &  0.90  \\ 
      6.0 & 0.05 & $10^{9.3}$ & 1.18 &  0.82 \\ 
      \hline
       7.0 & 0.41 & $10^{10.5}$ & 0.49 & 2.38 \\ 
       7.0 & 0.41 &$10^{9.8}$ & 0.37 & 1.81  \\ 
      7.0 & 0.41 & $10^{9.2}$ & 0.29 & 1.38 \\ 
      \hline
       8.0 & 0.67 & $10^{10.3}$ & 0.26 & 4.93  \\ 
       8.0 & 0.67& $10^{9.7}$ & 0.18 &  3.43 \\ 
      8.0 & 0.67 & $10^{9.2}$ & 0.12 & 2.25 \\ 
      \hline
     10.0 & 0.90 & $10^{9.8}$ & 0.10 &  15.55 \\ 
      10.0  & 0.90 & $10^{9.4}$ & 0.06 &  9.50 \\ 
      10.0 & 0.90 & $10^{9.1}$ & 0.04 & 5.40 \\ 
      \hline
    \end{tabular}
    \caption{Tabulation of the halo mean free path, $\Lambda_{\rm mfp}$, in the 40-2048 simulation for the five redshift/neutral fractions displayed in Fig.~\ref{fig:free_path_halo}.  The columns list, from left to right, the redshift, $z$, the volume weighted \HI fraction, $\langle x_{\rm HI} \rangle$, the average halo mass in each bin, $M_{\rm h}$, the halo mean free path, $\Lambda_{\rm mfp}$ and its ratio with the mean free path in the average IGM, $\lambda_{\rm mfp}$.}
    \label{tab:halo_mfp}
\end{table}

\begin{figure*}
\begin{centering}
    \includegraphics[width=0.48\textwidth]{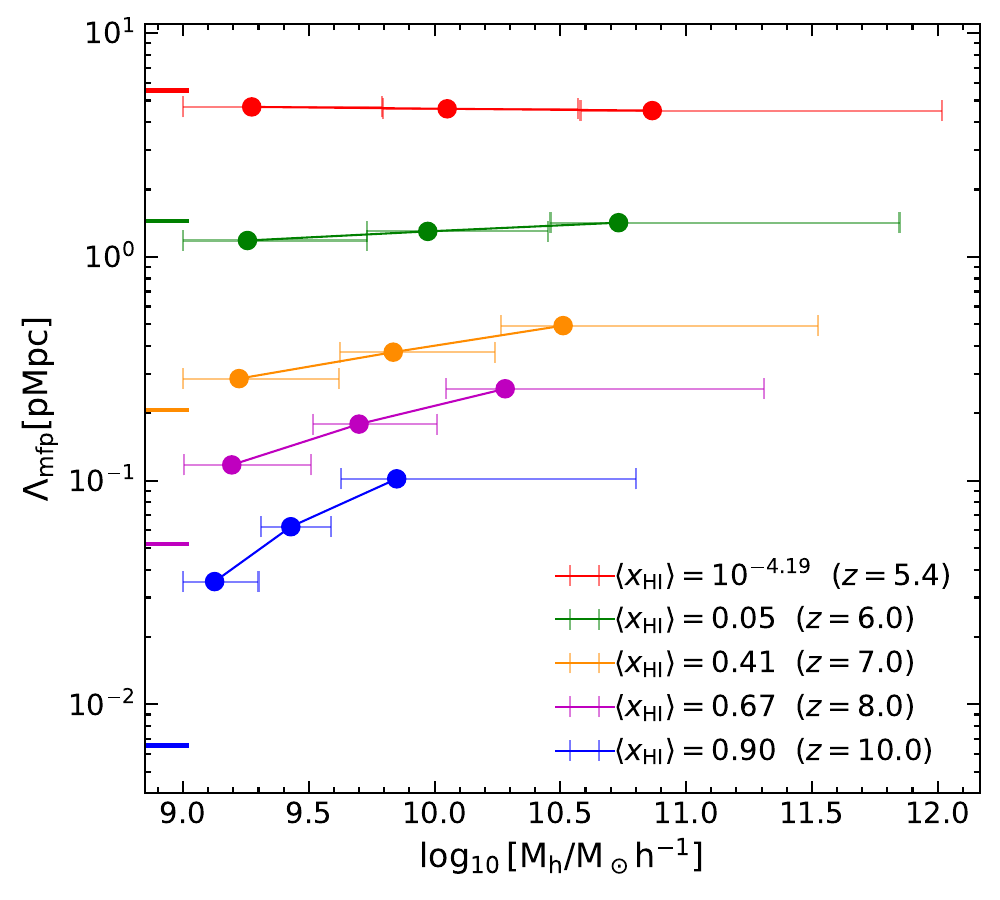}
       \includegraphics[width=0.48\textwidth]{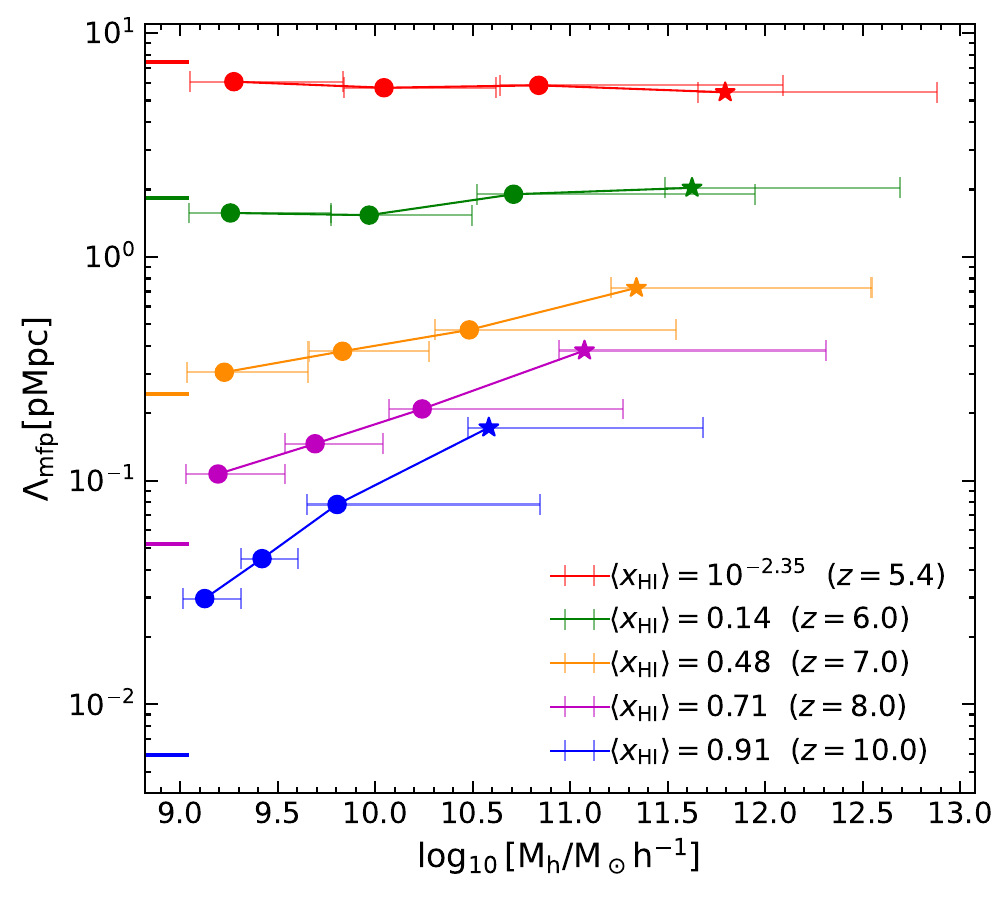}
    \vspace{-0.4cm}
    \caption{The Lyman-limit mean free path around haloes, $\Lambda_{\rm mfp}$, against the logarithm of the total halo mass, $\log_{10}(M_{\rm h}/h^{-1}\,{\rm M}_{\odot})$, for the five different neutral hydrogen fractions/redshifts used in Fig.~\ref{fig:pdf_over_a_range_of_z} (see also Table~\ref{tab:halo_mfp}).  The left panel shows the 40-2048 model, while the right panel shows the 160-2048 simulation.  The horizontal error bars give the size of each logarithmic mass bin (see text for details).  Note also the different scales on the horizontal axis of each panel, as well as the additional mass bin in the right panel (filled stars) for the $10^{3}$ most massive haloes in the 160-2048 simulation. The corresponding mean free path for the average IGM, matching the values displayed in Fig.~\ref{fig:overview_figure}, is shown by the horizontal lines on the left of each panel.}
    \label{fig:free_path_halo}
    \end{centering}
\end{figure*}

%%%%%%%%%%%%%%%%%%%%%%%%%%%%%%%%%%%%%%%%%%%%%%%%%%%%%%%%%%%%%%%%%%
%%%%%%%%%%%%%%%%%%%%%%%%%%% SECTION 4 %%%%%%%%%%%%%%%%%%%%%%%%%%%%
%%%%%%%%%%%%%%%%%%%%%%%%%%%%%%%%%%%%%%%%%%%%%%%%%%%%%%%%%%%%%%%%%%

\section{The mean free path around ionising source host haloes} \label{mfp_halos}
\subsection{Method for calculating the mean free path around haloes}

\begin{figure*}
\centering
  \includegraphics[width=0.95\textwidth]{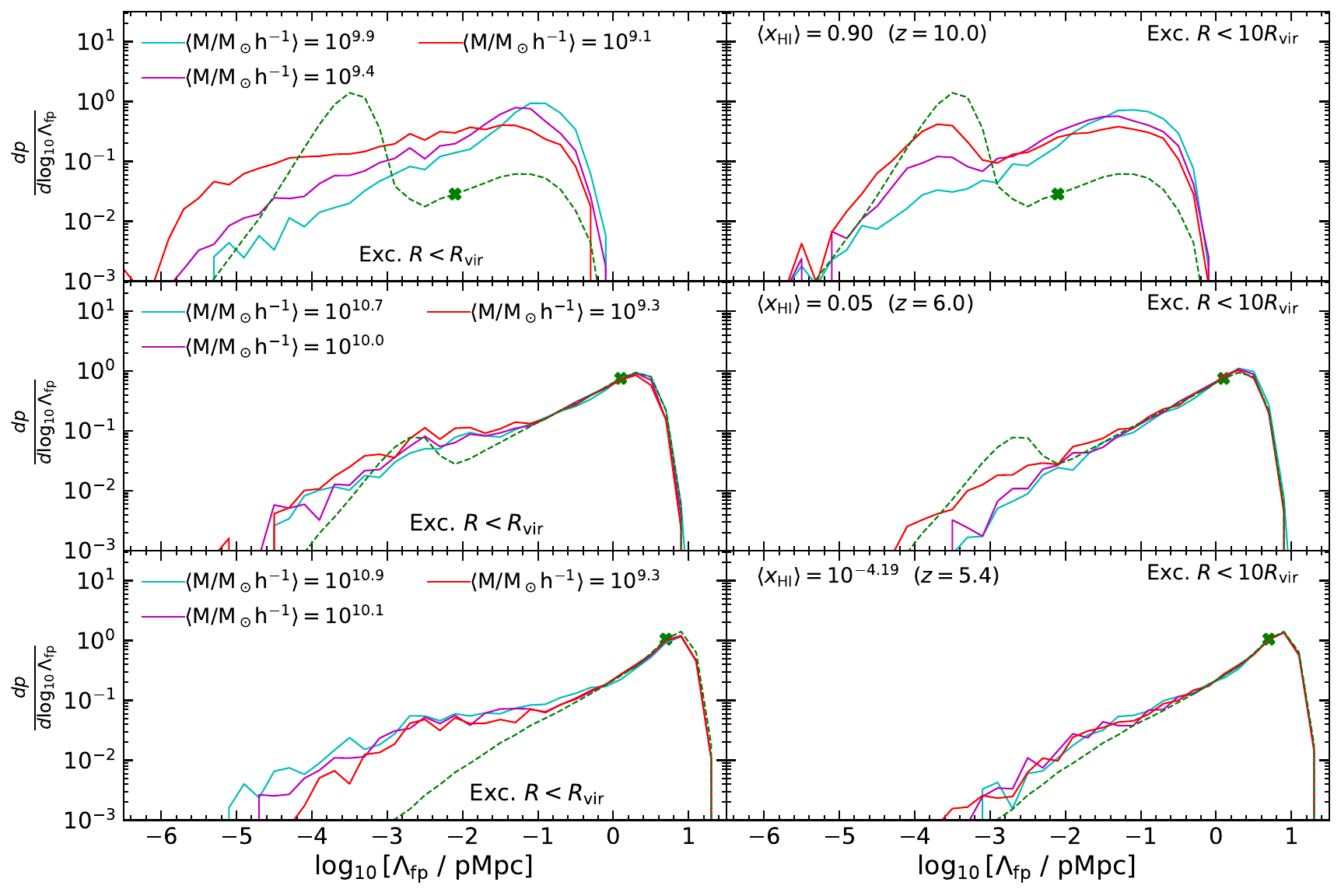
  }
  \vspace{-0.4cm}
  \caption{{\it Left column:} Probability density functions of the logarithm of the free path for Lyman-limit photons around haloes, $\log_{10}(\Lambda_{\rm fp}/\rm pMpc)$, in the 40-2048 simulation for three different volume weighted neutral hydrogen fractions/redshifts (the $\langle x_{\rm HI}\rangle$ values are indicated in the right column panels only).  Solid curves show $p(\log_{10}[\Lambda_{\rm fp}/\rm pMpc])$ for the three different mass bins indicated, while the dashed green curves show the corresponding average IGM distribution, $p(\log_{10}[\lambda_{\rm fp}/\rm pMpc])$. The mean free path for the average IGM, $\lambda_{\rm mfp}$, is shown as a cross.   All gas within one virial radius, $R_{\rm vir}$, of the centre of mass of the haloes has been excluded when calculating $\Lambda_{\rm fp}$ (see text for details).  {\it Right column:} As for the left column, but for free paths that now start at $10R_{\rm vir}$ from the halo centre of mass. Note these $\Lambda_{\rm fp}$ distributions fall more in line with the average IGM distribution (green dashed curves).} 
  \label{fig:visualising_halos}
\end{figure*}

\begin{figure*}
    \includegraphics[width=0.95\textwidth]{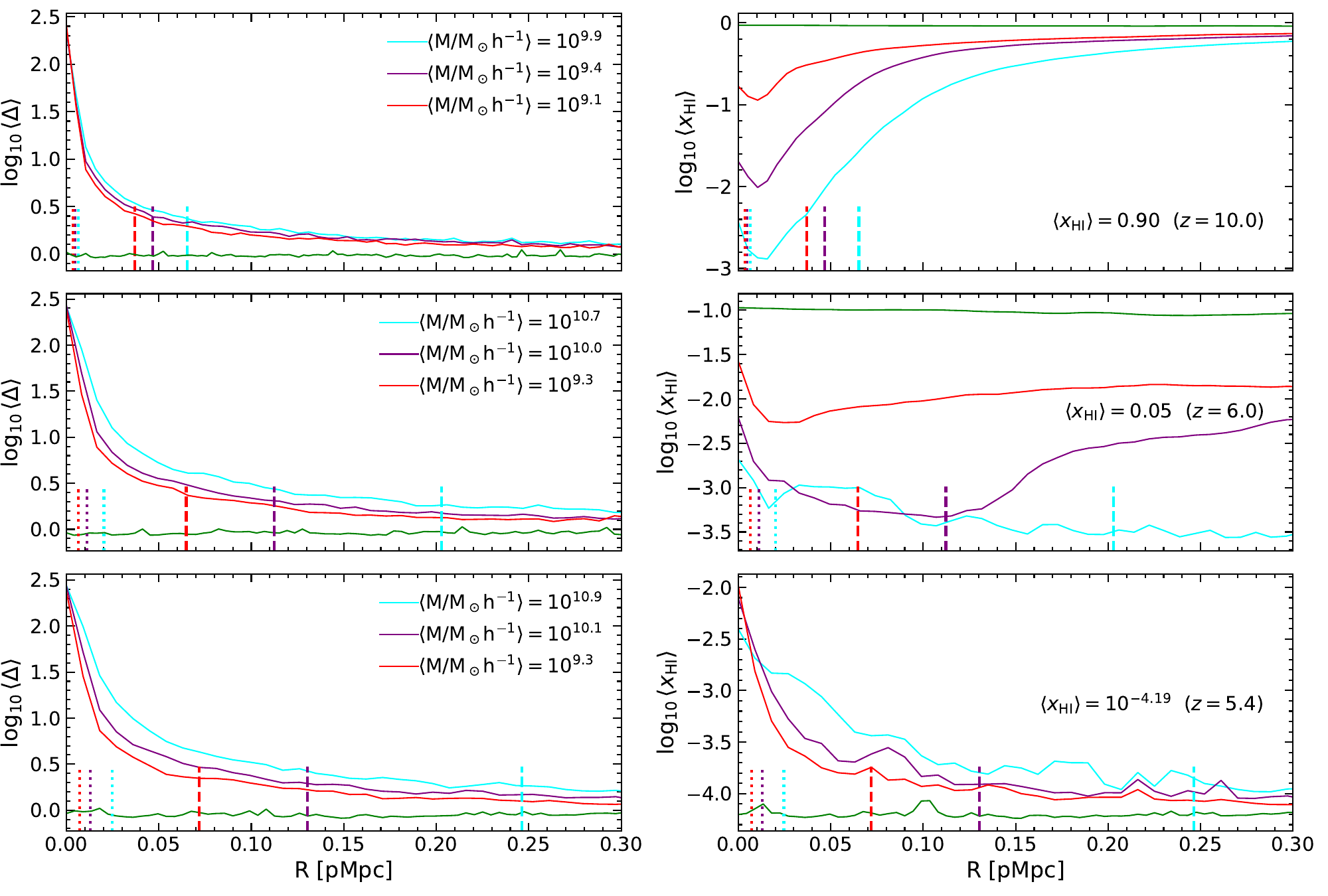}
    \vspace{-0.4cm}
    \caption{{\it Left column:} The logarithm of the average gas overdensity, $\log_{10} \langle \Delta \rangle$, against distance from the halo centre of mass in the 40-2048 simulation.  Each of the three halo mass bins may be compared to the average IGM (green curves).  From top to bottom, the results for three different volume weighted neutral fractions/redshifts are shown, $\langle x_{\rm HI} \rangle = 0.90,\,0.05$ and $10^{-4.19}$ (or $z=10,\,6$ and $5.4$). The dotted (dashed) vertical lines show the value of $R_{\rm vir}$ ($10R_{\rm vir}$) for the average halo mass in each bin (cf. Fig.~\ref{fig:visualising_halos}).  Larger overdensities are associated with the more massive haloes, as expected. {\it Right column}:  As for the left column, but now showing the logarithm of the average \HI fraction, $\log_{10} \langle x_{\rm HI} \rangle$.  \emph{Note the scale on the vertical axis is not the same in all panels.} }
    \label{fig:lines_of_sight}
\end{figure*}

Thus far we have investigated the Lyman-limit photon mean free path for the ``average'' IGM, yet $\lambda_{\rm mfp}$ measurements are typically determined from the Lyman continuum opacity in the vicinity of (stacked) high redshift quasars \citep{Prochaska_opacity_2009,becker_mean_2021,zhu_mfp_2023}.  The quasar proximity effect \citep{daloisio_large_2018}, neutral islands \citep{Roth_mfp_2023} or Lyman-limit systems around the quasar host haloes \citep{Prochaska2014} could then bias the mean free path measurements.  \citet{Satyavolu_mfp_2023} addressed these biases and concluded that -- if correctly accounted for -- they should not significantly impact the recent $\lambda_{\rm mfp}$ measurements at $5<z<6$. We do not revisit the issue of measurement bias here, therefore.  Instead, we focus on understanding the physical origin of variations in the mean free path around haloes in late reionisation models.

We first identify haloes in the simulations at the five neutral fractions/redshifts we have investigated previously.  At each redshift, we construct mass bins each containing $10^{3}$ unique haloes, where the smallest halo mass in the lower bin, $M=10^{9}h^{-1}\rm\,M_{\odot}$, corresponds to the minimum host halo mass for ionising sources in Sherwood-Relics \citep{puchwein_sherwood-relics_2022}.  The mean free path in each mass bin is then calculated in the same way as described in Section~\ref{mfp_calc}, except that (i) the sight-line now begins at a halo centre of mass, drawn in one of six directions (i.e., $\pm x,\pm y,\pm z$) parallel to the Cartesian axes of the simulation box,  and (ii) the Lyman-limit photon starting position is at one virial radius\footnote{At the redshifts we consider here, Eq.~(\ref{eq:virial}) is within $0.3$ per cent of the expression given by eq. (24) of \citet{BarkanaLoeb2001}.}, $R_{\rm vir}$, from the halo centre of mass, where 
\begin{align} 
R_{\rm vir} &\simeq \left(\frac{2G}{H_{0}^{2}\Omega_{\rm m}}\right)^{1/3}M_{\rm h}^{1/3}\Delta_{\rm c}^{-1/3}(1+z)^{-1},\nonumber\\ 
&= 8.7{\rm\,pkpc}\,\left(\frac{M_{\rm h}}{10^{10}\rm\,M_{\odot}}\right)^{1/3}\left(\frac{\Delta_{\rm c}}{18\pi^{2}}\right)^{-1/3}\left(\frac{1+z}{8}\right)^{-1}, \label{eq:virial}
\end{align}
as is appropriate for $\Omega_{\rm m}(1+z)^{3}\gg \Omega_{\Lambda}$.  This is because the Lyman-limit opacity within $R_{\rm vir}$ will depend on highly uncertain sub-grid physics that we do not model here (e.g., star formation and feedback, the properties of the interstellar medium and the escape of ionising radiation). 

Finally, to aid with distinguishing the halo mean free path from the mean free path for the average IGM, in the following discussion we introduce the notation, $\Lambda_{\rm mfp}$, for the Lyman-limit photon mean free path starting from a halo.

\subsection{Dependence of the mean free path on halo mass}

The halo mean free path, $\Lambda_{\rm mfp}$, against halo mass from the 40-2048 (left panel) and 160-2048 (right panel) simulations is displayed in Fig.~\ref{fig:free_path_halo} for five different \HI fractions/redshifts.  We also report the values for the 40-2048 model in Table~\ref{tab:halo_mfp}.  Note that, to facilitate comparison, the first three halo mass bins are identical for both simulations. The fourth mass bin in the right panel of Fig.~\ref{fig:free_path_halo}  (shown by the filled stars) contains the $10^{3}$ most massive haloes in the 160-2048 volume.

Several general trends are apparent in both models, despite the differences in the simulation mass resolution/box size and reionisation history.  First, regardless of the underlying \HI fraction, the halo mean free path is not equal to the average IGM value, $\Lambda_{\rm mfp}\neq \lambda_{\rm mfp}$, where the average IGM value, $\lambda_{\rm mfp}$, is indicated by the horizontal lines on the left of the panels in Fig.~\ref{fig:free_path_halo}.  Prior to the completion of reionisation the halo mean free path is greater than the average IGM value, $\Lambda_{\rm mfp}>\lambda_{\rm mfp}$, yet once the IGM is almost fully ionised this trend reverses, and $\Lambda_{\rm mfp}<\lambda_{\rm mfp}$.  The magnitude of this bias, relative to the average IGM value, $\lambda_{\rm mfp}$, changes with the volume averaged \HI fraction; at $z=5.4$, the halo mean free path, $\Lambda_{\rm mfp}$, is $\sim 20$ per cent smaller than the average IGM value, while at $z=10$, for the most massive haloes it is over an order of magnitude larger than the average IGM value.  Second, the dependence of $\Lambda_{\rm mfp}$ on halo mass, $M_{\rm h}$, varies with the \HI fraction.  During reionisation, $\Lambda_{\rm mfp}$ is correlated with halo mass, but after reionisation the trend is reversed, with halo masses that are (weakly) anti-correlated with $\Lambda_{\rm mfp}$.

The reason for this behaviour is explored further in the distribution of the halo free path, $\Lambda_{\rm fp}$, in Fig.~\ref{fig:visualising_halos}.  We show results for the 40-2048 model at $\langle x_{\rm HI} \rangle = 0.90,\, 0.05$ and $10^{-4.19}$  (corresponding to redshifts $z=10,\,6$ and $5.4$), for each of the mass bins.  For comparison, the dashed green curves show the distribution for the free path, $\lambda_{\rm fp}$, in the average IGM.   As before, the left hand column of Fig.~\ref{fig:visualising_halos} excludes all gas within $R_{\rm vir}$ of the halo centre of mass (i.e. as for Fig.~\ref{fig:free_path_halo}),  but the right column now excludes all gas within a larger region, $10R_{\rm vir}$.

We observe that -- at all stages of reionisation -- the smallest free paths are associated with gas between $1R_{\rm vir}$ and $10R_{\rm vir}$; these small free paths are removed from the distribution that excludes gas within $10R_{\rm vir}$, which converges toward the average IGM distribution.  The $\Lambda_{\rm fp}$ distribution is furthermore not as strongly bimodal as the average IGM.  At $\langle x_{\rm HI} \rangle=0.9$ ($z=10$), for the long mode of the average IGM distribution at $\lambda_{\rm fp}\sim 10^{-1}\rm\,pMpc$, the incidence of halo free paths is around an order of magnitude greater than for the average IGM.  However, following reionisation, when $\langle x_{\rm HI} \rangle=10^{-4.19}$ ($z=5.4$), there is instead an excess of halo free paths at the short mode for the average IGM, $\lambda_{\rm fp}\sim 10^{-3}\rm\,pMpc$.  As discussed in Section~\ref{systems_that_set_mfp}, these long (short) modes are typically associated with ionised (neutral) gas, indicating that haloes before (after) reionisation are preferentially probing the more ionised (neutral) hydrogen.  The smallest halo free path values at $z=5.4$ ($z=10$) are furthermore associated with the most (least) massive haloes, consistent with the behaviour in Fig.~\ref{fig:free_path_halo} where halo mass anti-correlates (correlates) with $\Lambda_{\rm mfp}$.  In other words, post-reionisation, differences between the mean free path in the IGM and around haloes are due to large scale structure bias, but when the IGM is more than a few per cent neutral by volume the differences instead arise from ionisation bias \citep[cf.][]{Wyithe_bias_2008}.

In Fig.~\ref{fig:lines_of_sight}, the role that the gas density and \HI fraction around the haloes play in setting this behaviour is highlighted.  The average gas overdensity (left column) and \HI fraction (right column) around haloes in the three mass bins used for the 40-2048 model are compared to the average IGM (green curves).  Each row shows the same three volume averaged \HI fractions/redshifts as Fig.~\ref{fig:visualising_halos}.  The more massive haloes are always associated with larger overdensities.  For a fixed \HI fraction, we would expect the mean free path to decrease with increasing halo mass, because $\Lambda_{\rm mfp}\propto (x_{\rm HI}\Delta)^{-1}$.  However, the \HI fraction around the haloes also varies as reionisation progresses.  During the early stages of reionisation, when $\langle x_{\rm HI} \rangle=0.9$, the \HI fraction around the most massive haloes is substantially \emph{smaller} than for lower mass haloes and the average IGM.  This is a consequence of (i) ionising source luminosity being proportional to halo mass in our model and (ii) the clustering of ionising sources around the most massive haloes (see Fig~\ref{fig:mfp_slice}).  This enhanced ionisation increases the mean free path relative to the average IGM, and causes $\Lambda_{\rm mfp}$ to correlate with halo mass, $M_{\rm h}$.

Similar behaviour is observed when the IGM is mostly reionised, with only a few neutral islands remaining (i.e., $\langle x_{\rm HI}\rangle = 0.05$ at $z=6$, in the middle row of Fig.~\ref{fig:lines_of_sight}).  Here the contrast in the \HI fraction with halo mass is smaller and the profiles vary less with distance from the halo centre of mass, $R$.  This is a consequence of the larger ionised regions around the haloes compared to higher redshift.  However, by $z=5.4$ the trend has reversed, and the most massive haloes have a larger average \HI fraction in their vicinity.  Here, the average photoionisation rate in the IGM, $\Gamma_{\rm HI}$, has increased by a factor of $2$--$3$ from $z=6$ (see Fig.~\ref{fig:overview_figure}), and the residual \HI in the IGM is now approximately in photoionisation equilibrium with an increasingly spatially uniform ionising background.  In this case, $x_{\rm HI}\propto \Delta$.  It is this interplay between gas density and spatial variations in the \HI fraction approaching haloes that drives the halo mean free path behaviour in Fig.~\ref{fig:free_path_halo}.

Our results are consistent with \citet{Rahmati_mfp_2018}, who reported a bimodal distribution for the free path that was halo mass dependent.  The halo mean free path was furthermore suppressed (enhanced) relative to the average IGM value following (during) reionisation, matching our findings.  Our results are also broadly similar to \citet{Fan2024}, who reported a slightly smaller suppression in the mean free path around $\sim 10^{11}$--$10^{12}\rm\,M_{\odot}$ haloes of $\sim 10$ per cent at $z\sim 5$. In the analytical halo model for Lyman-limit systems presented by \citet{Theuns_LLS_2024} the mean free path was also smaller around the massive haloes hosting quasars at $z\simeq 6$ when compared to the average IGM \citep[see also][]{Prochaska2014}. \citet{Theuns_LLS_2024} attributed this difference to the clustering of Lyman-limit systems around the quasar host haloes; the ratio between the attenuation length around haloes and the average IGM in their model was around an order of magnitude for the halo masses, $M_{\rm h}=10^{12}$--$10^{13}\rm\,M_{\odot}$, thought to be typical of quasar hosts \citep{Pizzati2024}.  Our 40-2048 simulation only includes haloes up to  $10^{12}\rm\,M_{\odot}$ at $5<z<6$, but do we find a small reduction, $0.8 \lesssim \Lambda_{\rm mfp}/\lambda_{\rm mfp}\lesssim 1.0$, at the end of reionisation (Table~\ref{tab:halo_mfp}).  In the 160-2048 model, which includes haloes with masses up to $M_{\rm h}=10^{12.8}\rm\,M_{\odot}$ at $z=6$, we instead find $\Lambda_{\rm mfp}/\lambda_{\rm mfp}\simeq 0.7$ for the most massive haloes, although note again that we ignore all gas within the virial radius of the haloes.  Our results suggest the \citet{Theuns_LLS_2024} model will not hold during reionisation, when the mean free path is no longer set by neutral gas in high density, small-scale structures.

In summary, any systematic biases should be accounted for when measuring the mean free path around bright quasars at $z>5$, but as pointed out in recent work \citep{Satyavolu_mfp_2023,Roth_mfp_2023} these already appear to be well controlled in the most recent measurements.  The biases we discuss here will furthermore be sub-dominant to effect of the significantly enhanced ionisation around bright quasars at $z\simeq 6$, which we do not model in this work.  Lastly, as has been noted elsewhere \citep[e.g.][]{Mason2018,Neyer2023,Lu2024,Asthana2024}, a mean free path that is significantly larger than in the average IGM around ionising source host haloes during reionisation will aid the visibility of high redshift \Lya emitting galaxies at $z>7$.  Indeed, multiple objects have now been detected with near-infrared spectroscopic observations \citep[e.g.][]{Larson2022,Tang2023,Whitler2024}, with the current highest redshift \Lya emitter recently detected at $z=10.6$ with JWST \citep{Bunker2023}.

%%%%%%%%%%%%%%%%%%%%%%%%%%%%%%%%%%%%%%%%%%%%%%%%%%%%%%%%%%%%%%%%%%
%%%%%%%%%%%%%%%%%%%%%%%%%%% SECTION 5 %%%%%%%%%%%%%%%%%%%%%%%%%%%%
%%%%%%%%%%%%%%%%%%%%%%%%%%%%%%%%%%%%%%%%%%%%%%%%%%%%%%%%%%%%%%%%%%

\section{Conclusions} \label{conclusions}

In this work we have examined the mean free path to Lyman-limit photons, $\lambda_{\rm mfp}$, predicted by the Sherwood-Relics simulation suite \citep{puchwein_sherwood-relics_2022}.  Sherwood-Relics has been specifically designed for modelling the \Lya forest at the end stages of inhomogeneous reionisation.  Here we consider a sub-set of the Sherwood-Relics simulations that have been calibrated to reproduce the observed \Lya forest transmission at $5<z<6$, assuming a late end to reionisation (i.e. at $z<6$).  The simulations self-consistently follow the hydrodynamical response of the diffuse IGM to photoionisation and heating, and our fiducial model has a mass resolution intermediate between the recent \textsc{thesan} \citep{Garaldi_THESAN_2022} and CoDa III  \citep{lewis_short_2022} radiation hydrodynamical simulations.  Importantly for this study, while Sherwood-Relics does not explicitly model star formation and dense, cool gas with $\Delta = \rho/\langle \rho \rangle >1000$ and $T<10^{5}\rm\,K$, it follows the absorption systems in the \Lya forest that dominate the Lyman-limit opacity during the final stages of late reionisation \citep{Rahmati_mfp_2018,Nasir_photon_sink_2021}.  Our main findings are as follows:

\begin{itemize}

\item The Lyman-limit mean free path predicted by our fiducial Sherwood-Relics simulation (40-2048) is reasonably well converged (within 5 per cent) with mass resolution and is in good agreement  (typically within $1\sigma$) with recent measurements of the mean free path from direct determinations using the Lyman continuum opacity \citep{becker_mean_2021,zhu_mfp_2023} and indirect determinations using the observed \Lya forest transmission \citep{gaikwad_measuring_2023}.  At $z=5.93$, the 40-2048 simulation furthermore predicts a mean free path $1.2\sigma$ above the recent measurement presented by \citet{zhu_mfp_2023}, although the mean reionisation history in this model will not be converged with box size.  Our larger simulation volume at lower resolution (160-2048) is $1.8\sigma$ above the same measurement and has a reionisation history that should be marginally converged with box size \citep[cf.][]{Iliev2014}. However, the lower mass resolution of this model will overpredict the mean free path at fixed \HI fraction by at least $20$ per cent. \\

\item In order to match the observed \Lya transmission, reionisation completes at $z\sim 5.7$ in our fiducial model, and at $z\sim 5.3$ in our larger, lower resolution volume.  In both cases, the observed evolution of $\lambda_{\rm mfp}$ is consistent with the emerging consensus that reionisation ends at $5<z<6$ \citep[see also][]{keating_constraining_2020,lewis_short_2022,Garaldi_THESAN_2022,bosman_hydrogen_2022}. \\

\item The Lyman-limit photon free path distribution in Sherwood-Relics is bimodal during reionisation, with a minimum at $\lambda_{\rm fp}\sim 10^{-2}\rm\,pMpc$.  The short (long) mode is typically associated with neutral (ionised) hydrogen, and is a generic prediction of inhomogeneous reionisation models \citep[cf.][]{lewis_short_2022,Satyavolu_mfp_2023}.  This bimodality also manifests in the \HI column density distribution, with \HI column densities at $N_{\rm HI}\sim 10^{19}\rm\,cm^{-2}$ ($N_{\rm HI}\sim 10^{15.5}\rm\,cm^{-2}$) associated with the neutral (ionised) hydrogen in the diffuse IGM.\\

\item During the final stages of late reionisation at $z=6$, when the volume averaged \HI fraction is $\langle x_{\rm HI} \rangle=0.05$ in our fiducial model, the majority of the total Lyman-limit opacity (61 per cent) arises from the residual \HI fraction in the \emph{already reionised intergalactic medium}.  The residual \HI is typically associated with absorber column densities $N_{\rm HI}=10^{16}$--$10^{17}\rm\,cm^{-2}$ with neutral fractions $x_{\rm HI}\sim 10^{-4}$--$10^{-2}$, in broad agreement with earlier work by \citet{Rahmati_mfp_2018} using radiation hydrodynamical simulations performed in smaller volumes and at lower mass resolution.\\

\item As a consequence of the typical \HI absorption systems, $N_{\rm HI}\sim 10^{16}$--$10^{17}\rm\,cm^{-2}$, that dominate the Lyman-limit opacity at the tail-end of reionisation, we argue that an important requirement for inhomogeneous reionisation simulations that capture the redshift evolution of the mean free path at the end of reionisation is resolving photon sinks in the diffuse IGM at moderate overdensities.  Capturing these systems will also require accurate modelling of the hydrodynamical response of the gas to photo-heating during reionisation \citep[e.g.][]{puchwein_sherwood-relics_2022}.\\

\item The mean free path around dark matter haloes, $\Lambda_{\rm mfp}$, differs from the average IGM throughout and after reionisation.  We consider halo masses in the range $M_{\rm h}\sim 10^{9}$--$10^{11}\,h^{-1}{\rm M}_{\odot}$ in our fiducial simulation, but we observe the same general trends in a lower resolution simulation with 64 times the volume.  Throughout most of reionisation the halo mean free path is greater than the average IGM value, but  during the final stages of reionisation it is smaller.  The enhancement during reionisation (by as much as an order of magnitude) is because of the highly ionised bubbles centred around (clustered) haloes; these will be relevant for aiding the visibility of high redshift \Lya emitters at $z>7$ \citep[e.g.][]{Larson2022,Tang2023,Whitler2024,Asthana2024}.  The suppression of $\Lambda_{\rm mfp}$ following reionisation (typically by $\sim 20$ per cent relative to the average IGM) is because of the higher density of gas around haloes once the IGM is approximately in photoionisation equilibrium, such that $x_{\rm HI} \propto \Delta$.\\

\item During reionisation, the halo mean free path correlates with halo mass, partly because of ionising source clustering but also because our modelling assumes the source ionising luminosity is proportional to halo mass.  However, following reionisation, the halo mean free path is weakly anti-correlated with halo mass \citep[see also][]{Theuns_LLS_2024}.  These biases should be accounted for when measuring the mean free path around bright quasars at $z>5$, but by $z=6$ they are likely to be sub-dominant to the significantly enhanced ionisation from the proximity effect around bright quasars \citep[see e.g.][]{Satyavolu_mfp_2023,Roth_mfp_2023}. 

\end{itemize}

\noindent
Further progress with modelling the Lyman-limit mean free path at the end of reionisation relies on capturing a large dynamic range.  Our results suggest that a $160h^{-1}\rm\,cMpc$ box with a dark matter particle mass of $5.37\times 10^{5}h^{-1}\,M_{\odot}$ (or equivalently $2\times 8192^{3}$ particles) should yield a mean free path that is reasonably well converged with both box size and mass resolution.  The required number of resolution elements is already similar to the CoDa III simulation \citep{lewis_short_2022}, but at present such models are expensive and have data volumes that are challenging to handle.  Accounting for potential biases in mean free path measurements may also be valuable, although recent work has suggested these should already be well controlled in the existing data at $z\simeq 6$ \citep{Satyavolu_mfp_2023,Roth_mfp_2023}.  Finally, the rather low emissivity preferred by the \Lya forest and Lyman-limit mean free path at $5<z<6$ (where $\dot{n}=3\times 10^{50}\rm\,s^{-1}\,cMpc^{-3}$ corresponds to only $1$--$2$ ionising photons emitted per hydrogen atom over the age of the Universe at $z=6$) remains an important constraint on the reionisation photon budget, particularly given the apparent abundance of ionising photons at $z>6$ now implied by JWST \citep{Munoz2024}.  This tension may necessitate substantial redshift evolution in the properties of ionising photon sources at $z>6$, or a revision in our understanding of the IGM and structure formation during the reionisation era.

\section*{Acknowledgements}
The simulations used in this work were performed using the Joliot Curie supercomputer at the Tr{\'e}s Grand Centre de Calcul (TGCC) and the Cambridge Service for Data Driven Discovery (CSD3), part of which is operated by the University of Cambridge Research Computing on behalf of the STFC DiRAC HPC Facility (www.dirac.ac.uk).  We acknowledge the Partnership for Advanced Computing in Europe (PRACE) for awarding us time on Joliot Curie in the 16th call. The DiRAC component of CSD3 was funded by BEIS capital funding via STFC capital grants ST/P002307/1 and ST/R002452/1 and STFC operations grant ST/R00689X/1.  This work also used the DiRAC@Durham facility managed by the Institute for Computational Cosmology on behalf of the STFC DiRAC HPC Facility. The equipment was funded by BEIS capital funding via STFC capital grants ST/P002293/1 and ST/R002371/1, Durham University and STFC operations grant ST/R000832/1. DiRAC is part of the National e-Infrastructure.  JSB, LC and EC are supported by STFC consolidated grant ST/X000982/1. EC also acknowledges the support of a Royal Society Dorothy Hodgkin Fellowship and a Royal Society Enhancement Award. Support by ERC Advanced Grant 320596 ‘The Emergence of Structure During the Epoch of Reionization’ is gratefully acknowledged. MGH has been supported by STFC consolidated grants ST/N000927/1 and ST/S000623/1.  We thank Volker Springel for making P-Gadget-3 available. We also thank Dominique Aubert for sharing the ATON code, and Philip Parry for technical support. For the purpose of open access, the author has applied a Creative Commons Attribution (CC BY) licence to any Author Accepted Manuscript version arising from this submission.

%%%%%%%%%%%%%%%%%%%%%%%%%%%%%%%%%%%%%%%%%%%%%%%%%%
\section*{Data Availability}
All data and analysis code used in this work are available from the first author on reasonable request.  Further guidance on accessing the publicly available Sherwood-Relics simulation data may also be found at \url{https://www.nottingham.ac.uk/astronomy/sherwood-relics/}.



\bibliographystyle{mnras}
\bibliography{refs} % if your bibtex file is called example.bib

\appendix

%%%%%%%%%%%%%%%%%%%%%%%%%%%%%%%%%%%%%%%%%%%%%%%%%%%%%%%%%%%%%%%%%%
%%%%%%%%%%%%%%%%%%%%%%%%%% APPENDIX A %%%%%%%%%%%%%%%%%%%%%%%%%%%%
%%%%%%%%%%%%%%%%%%%%%%%%%%%%%%%%%%%%%%%%%%%%%%%%%%%%%%%%%%%%%%%%%%

\section{Effect of the quick Lyman-$\alpha$ approximation on the mean free path} \label{app:qLya}

\begin{figure}
    \includegraphics[width=0.47\textwidth]{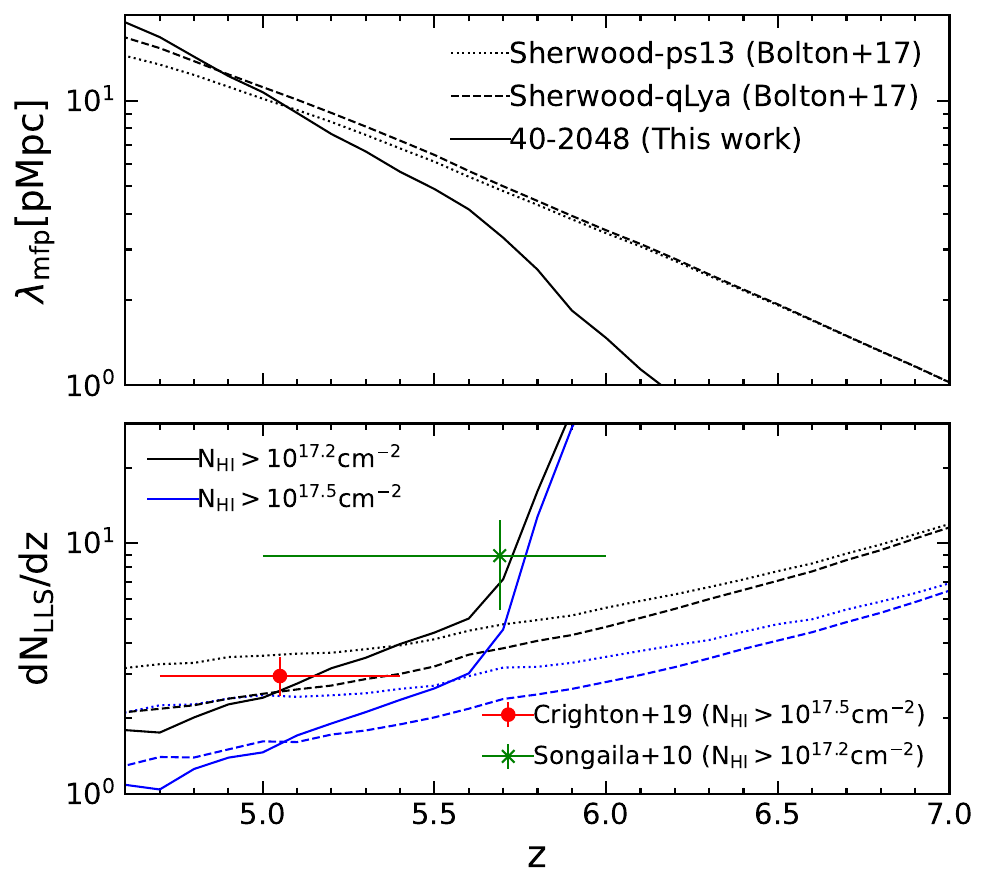}
    \vspace{-0.3cm}
    \caption{\emph{Top panel:} Comparison of the mean free path in two of the Sherwood simulations \citep{Bolton2017} to the 40-2048 simulation (solid curve) used in this work.  Sherwood-ps13 (dotted curve) uses a sub-grid model for star formation and feedback \citep{Puchwein2013} and explicitly follows high density gas in haloes, whereas Sherwood-qLy$\alpha$ (dashed curve) uses the same approximation as 40-2048, where all gas with $\Delta>1000$ and $T<10^{5}\rm\,K$ is converted into collisionless particles.  Both Sherwood models assume a homogeneous UV background, following \citet{HaardtMadau2012}, and the IGM remains highly ionised at $z>6$.  \emph{Bottom panel:}  The corresponding number of Lyman limit systems per unit redshift for $N_{\rm HI}>10^{17.2}\rm\,cm^{-2}$ (black curves) and $N_{\rm HI}>10^{17.5}\rm\,cm^{-2}$ (blue curves).  Observational data are from \citet{Songaila_Cowie_2010} (green cross) and \citet{Crighton2019} (red diamond).}
    \label{fig:dndz}
\end{figure}

In this Appendix we test the effect of the quick-Ly$\alpha$ approximation used in Sherwood-Relics on the Lyman-limit mean free path at $z>5$.  Although we do not have a Sherwood-Relics model that includes gas with $\Delta>1000$ and $T<10^{5}\rm\,K$, we may gain insight using two simulations from the original Sherwood simulation project \citep{Bolton2017}.  Here we make use of the 40-1024 and 40-1024-ps13 models presented in table 1 of \citet{Bolton2017}, which we refer to here as Sherwood-qLy$\alpha$ and Sherwood-ps13, respectively.  These have the same box size as the 40-2048 model, but use a dark matter particle mass that is $8$ times larger.  The Sherwood models do not model inhomogeneous reionisation.  Instead, a spatially uniform UV background with an IGM that remains highly ionised at $z>6$ is assumed \citep{HaardtMadau2012}. A correction for the self-shielding of dense hydrogen gas to ionising photons is also included, following \citet{Chardin2018}.  For the purpose of this comparison, the key difference is that Sherwood-qLy$\alpha$ uses the same quick-Ly$\alpha$ approximation used in this work, whereas Sherwood-ps13 uses the star formation and energy-driven outflow model from \citet{Puchwein2013}.  The models are otherwise identical; the difference between them gives an indication of how missing gas with $\Delta>1000$ and $T<10^{5}\rm\,K$ will alter the predicted mean free path.

The results of the comparison are displayed in Fig~\ref{fig:dndz}.  The upper panel shows the mean free path, while the lower panel displays the number of Lyman-limit systems per unit redshift, $dN_{\rm LSS}/dz$.  The difference in the mean free path predicted by Sherwood-ps13 and Sherwood-qLy$\alpha$ is at most 9 per cent $z\geq 5$, with an average difference of $3$ per cent at $5\leq z \leq 7$.  This difference decreases with increasing redshift as the typical overdensity of gas that is optically thick to Lyman-limit photons decreases.  A similar trend is observed for $dN_{\rm LSS}/dz$, where the difference between the Sherwood models at $z\simeq 5$ is comparable to the uncertainty on the observational measurements from \citet{Crighton2019}.  Note, however, that the effect of missing gas will be much larger at lower redshifts as the typical overdensity of Lyman-limit systems increases.  At $z<3$, the quick-Ly$\alpha$ model does not reproduce the observed incidence of Lyman limit systems.\footnote{See, for example, figure 1 in \citet{Miller2019}.}   

We furthermore note that the 40-2048 model is already in good agreement with the observed Lyman-limit systems with $N_{\rm HI}>10^{17.2}\rm\,cm^{-2}$ at $z=6$, although it underpredicts the number of systems with $N_{\rm HI}>10^{17.5}\rm\,cm^{-2}$ at $z=5$.  As discussed in Section~\ref{systems_that_set_mfp} and Appendix~\ref{sec:window_size} below, however, we caution that the number of systems will be sensitive to our choice of integration window size.

%%%%%%%%%%%%%%%%%%%%%%%%%%%%%%%%%%%%%%%%%%%%%%%%%%%%%%%%%%%%%%%%%%
%%%%%%%%%%%%%%%%%%%%%%%%%% APPENDIX B %%%%%%%%%%%%%%%%%%%%%%%%%%%%
%%%%%%%%%%%%%%%%%%%%%%%%%%%%%%%%%%%%%%%%%%%%%%%%%%%%%%%%%%%%%%%%%%

\section{Choice of integration window for calculating column densities}
\label{sec:window_size} 

As discussed in Section~\ref{systems_that_set_mfp}, we must select a length scale for computing \HI column densities, $N_{\rm HI}$.  In this work we have considered windows equal to the (redshift dependent) Jeans length at the mean background density \citep{schaye2001, Rahmati_mfp_2018}, and a fixed window of $50\rm\, km\,s^{-1}$ \cite[e.g.,][]{Gurvich2017}.  We obtain the proper Jeans length, $\lambda_{\rm J}$, for gas at the mean density using,
\begin{equation}
     \lambda_{\rm J} = \left(\frac{40\pi^2k_{\rm B}\langle T \rangle }{9\langle \mu \rangle m_{\rm H} H^2_0 \Omega_{\rm m} (1+z)^{3}}\right)^{1/2}.
     \label{eq:jeans}
 \end{equation}
Here $\langle T \rangle $ is the volume averaged temperature in the simulation, and $\langle \mu \rangle $ is the volume averaged mean molecular weight,
\begin{equation} 
\label{eq:mu}
\langle \mu \rangle = \mu_{\rm ionised}(1-f_{\rm neutral})+ \mu_{\rm neutral}f_{\rm neutral}, 
\end{equation}
where $\mu_{\rm ionised}=0.61$ for an \HII and \HeII admixture, $\mu_{\rm neutral}=1.22$ and $f_{\rm neutral}$ is the fraction of the simulation volume with $x_{\rm HI}>0.5$.  Adding some representative values, Eq.~(\ref{eq:jeans}) then reduces to,
\begin{equation}
  \lambda_{\rm J}=90.5\,{\rm\,pkpc}\,\left(\frac{\langle T \rangle}{10^{4}\rm\,K}\right)^{1/2}\left(\frac{\langle \mu \rangle}{0.61}\right)^{-1/2}\left(\frac{1+z}{8}\right)^{-3/2}.
    \label{eq:jeans-mean}
\end{equation}

\begin{figure}
    \includegraphics[width=0.45\textwidth]{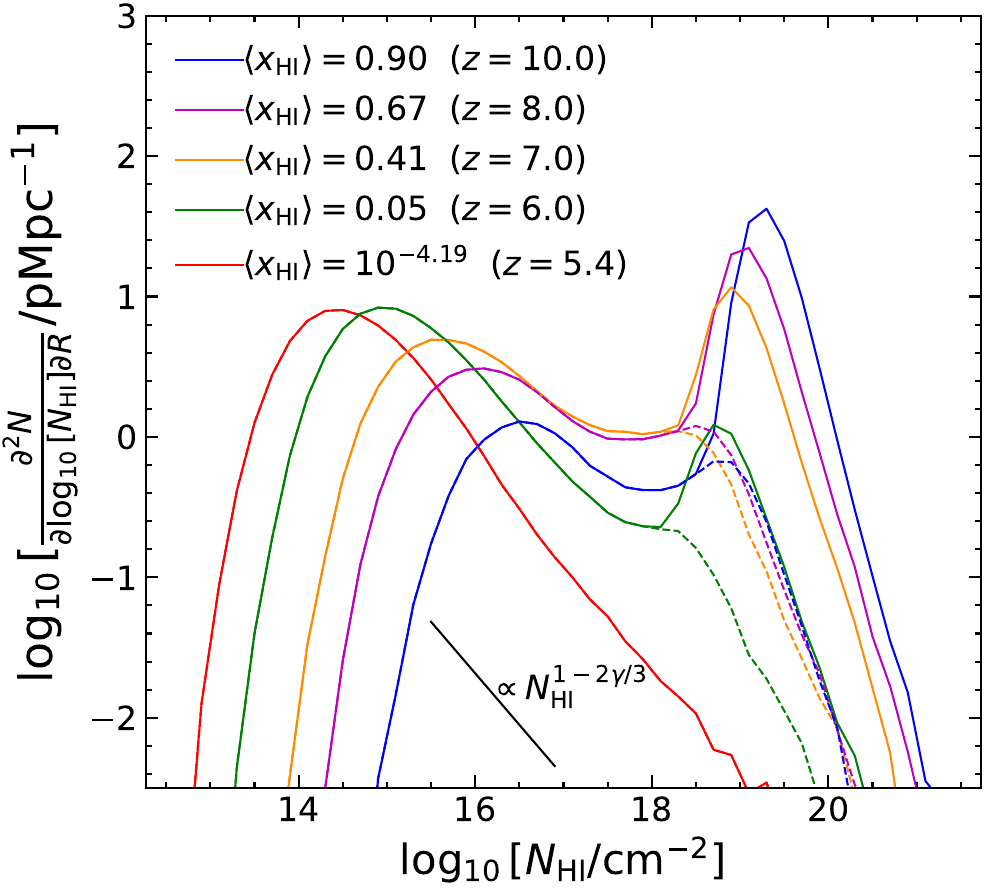}
    \vspace{-0.3cm}
    \caption{The \HI column density distribution, as described in Fig.~\protect\ref{fig:column_density_threhold}, but now $N_{\rm HI}$ has been calculated by integrating over a fixed window of size $50\rm \,km\,s^{-1}$. Note that although the peaks of the distribution appear at different column densities, the shape of the distribution is robust to the window choice.}
    \label{fig:50_window}
\end{figure}

\noindent
Fig.\ref{fig:column_density_threhold} displays the column density distribution function obtained using Eq.~(\ref{eq:jeans}). For comparison, Fig.~\ref{fig:50_window} shows the column density distribution for when fixed windows of $50\rm\, km\,s^{-1}$ are used.  The overall shape of the distribution is consistent between the two window sizes; we do not expect qualitative differences to arise from our choice of integration window.  It should nevertheless be noted that the absolute values for $N_{\rm HI}$ will always depend on the choice of integration window. 

%%%%%%%%%%%%%%%%%%%%%%%%%%%%%%%%%%%%%%%%%%%%%%%%%%%%%%%%%%%%%%%%%%
%%%%%%%%%%%%%%%%%%%%%%%%%% APPENDIX C %%%%%%%%%%%%%%%%%%%%%%%%%%%%
%%%%%%%%%%%%%%%%%%%%%%%%%%%%%%%%%%%%%%%%%%%%%%%%%%%%%%%%%%%%%%%%%%

\section{Analytical scaling relations}
\label{sec:scaling}

A commonly used approximation for modelling the Lyman-limit photon mean free path and the opacity of the IGM -- particularly when correcting for small-scale, unresolved photon sinks in the already reionised IGM -- is to use an analytical relation that captures how the opacity varies with the local \HI photoionisation rate \citep[e.g.][]{mcquinn2011,Davies2016,becker_mean_2021}.  The details of this scaling relation derive from the model presented in \citet[][hereafter MHR00]{miralda-escude2000}, as already discussed elsewhere \citep{furlanetto2005,mcquinn2011}.  Here we find it useful to recapitulate this model for a direct comparison to Sherwood-Relics.

In the \mhr{} analytical reionisation model, all gas up to some critical density, $\Delta_{\rm i}$ is assumed ionised, and gas above this density is fully neutral.   The mean free path of ionising photons is then proportional to the chance of an ionising photon meeting a neutral photon sink.  Given the volume-filling fraction of ionised gas $F_{\rm V}$, \mhr{} make the ansatz that the typical extent of a ionising photon sink is $(1-F_{\rm V})^{1/3}$; the chance of encountering a photon sink along a random line of sight is then $(1-F_{\rm V})^{1/3}/(1-F_{\rm V}) = (1-F_{\rm V})^{-2/3}$.  Hence, the mean free path $\lambda \propto (1-F_{\rm V})^{-2/3}$. 

It is straightforward to compute $F_{\rm V}$ given the differential probability distribution function for the gas density, $P(\Delta)$, 
\begin{equation}
  \label{eq:fv} F_{\rm V} = \int_0^{\Delta_{\rm i}}P(\Delta) \ud \Delta,
\end{equation} and so
\begin{equation}
  \label{eq:lam} \lambda \propto \left[ \int_{\Delta_i}^{\infty}P(\Delta) \ud \Delta \right]^{-2/3} .
\end{equation}

\noindent
\mhr{} showed that $P(\Delta)$ from cosmological hydrodynamical simulations \citep{Miralda1996} is well fit over the range $2\leq z \leq 4$ using a lognormal distribution with a high density, power-law tail,

\begin{equation}
  \label{eq:pdelta} P(\Delta)\ud \Delta = A\, {\rm exp}\left[ - \frac{\left(\Delta^{-2/3} - C_0 \right)^2}{2 \left( \delta_0/3 \right)^2} \right] \Delta^{-\gamma} \ud \Delta. 
\end{equation} 
\mhr{} extrapolated this fit to higher redshifts, $z>4$, assuming $\gamma = 5/2$.

Turning now to the production and absorption of ionising photons, assuming that the mean free path of an ionising photon is  much smaller than the horizon scale, such that cosmological effects can be neglected \citep[sometimes called the ``local source'' approximation, e.g.][]{Madau1999,schirber2003}, then we can write $\Gamma \propto \epsilon \lambda$, where $\Gamma$ is the hydrogen photoionisation rate and $\epsilon$ is the ionising emissivity (i.e. the number of ionising photons emitted per unit time per unit volume).  When the IGM is highly ionised, the emissivity is approximately balanced by the recombination rate, $\epsilon \simeq \mathcal{R}$, where
\begin{equation}
  \label{eq:rec-gen} \mathcal{R} = \alpha(T)n_{\rm e}n_{\rm HII}.
\end{equation}
Here $\alpha(T)$ is the recombination coefficient, $n_{\rm e}$ is the number density of electrons and $n_{\rm HII}$ is the number density of ionised hydrogen. If the gas is approximately isothermal, $\alpha(T)\simeq \alpha$, and if it is also highly ionised, $n_{\rm e} \approx n_{\rm HII}$.  We may then write
\begin{align}
  \label{eq:rec-app} \mathcal{R} = \alpha \bar{n}_{\rm e}^2 \int_{0}^{\Delta_i} \Delta^2 P(\Delta)\ud \Delta,
\end{align} where $\bar{n}_{\rm e}$ is the mean number density of electrons.

The final step is to notice that most recombinations happen at high densities, and  that Eq.~(\ref{eq:pdelta}) tends to a power law $P(\Delta)\propto \Delta^{-\gamma}$ for $\Delta \gg 1$. Plugging this into Eq.~(\ref{eq:lam}) we find
\begin{equation}
  \label{eq:lam-del} \lambda \propto \Delta_i^{2(\gamma-1)/3},
\end{equation} and into \eref{eq:rec-app} yields
\begin{equation}
  \label{eq:rec-del} \mathcal{R} \propto \Delta_i^{3-\gamma}.
\end{equation}  Assuming $\Gamma \propto \epsilon \lambda \simeq \mathcal{R} \lambda$ and combining equations~\eqref{eq:lam-del} and \eqref{eq:rec-del} gives an expression for the photoionisation rate
\begin{equation}
  \label{eq:gam-del} \Gamma \propto \Delta_i^{(7-\gamma)/3}.
\end{equation}
We can now use \eqref{eq:gam-del} in \eqref{eq:lam-del} to remove the explicit dependence on $\Delta_{\rm i}$, giving
\begin{equation}
  \label{eq:gam-xi} \lambda\propto \Gamma^\xi,
\end{equation}
where $\xi={2(\gamma-1)}/{(7-\gamma)}$. For the value of $\gamma=5/2$ presented in \mhr, this leads to  $\lambda\propto\Gamma^{2/3}$.  For the opacity, $\kappa \propto \lambda^{-1}$, and so $\kappa \propto \Gamma^{-2/3}$ \citep[e.g.,][]{Davies2016,becker_mean_2021}.

\begin{figure} 
    \centering
    \includegraphics[width=\linewidth]{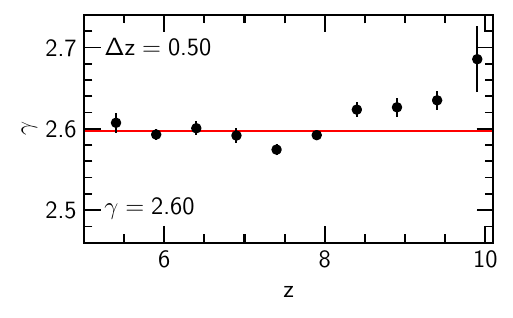}
    \vspace{-0.6cm}
    \caption{Estimate of the slope of the power-law high density tail, $P(\Delta)\propto\Delta^{-\gamma}$, in \eref{eq:pdelta} from the 40-2048 Sherwood-Relics model used in this work.  The values of $\gamma$ and their associated $1\sigma$ standard deviation are obtained by fitting $p(\Delta)$ in the simulation in steps of $\Delta z=0.1$, before combining into bins of $\Delta z=0.5$ to reduce noise (black points). The red line indicates a constant value of $\gamma = 2.6$ provides a reasonable fit to the data at $z<8$.  \label{fig:gam-fit}}
\end{figure}

We first check the expected slope of the high density power-law tail, $\gamma$, in \eref{eq:pdelta}. In \fref{fig:gam-fit} we show the result of fitting $P(\Delta)\propto \Delta^{-\gamma}$ for ${\rm log}_{10}\Delta \geq 1.2$ from $z=5.2$ to $z=10.1$ in the 40-2048 model. We group the individual values of $\gamma$ into bins of width $\Delta z = 0.5$ to reduce noise -- within each bin we compute the mean $\gamma$ and standard deviation.  We find that $\gamma$ is not independent with redshift, but a value of $\gamma = 2.6$ provides a reasonable approximation during the second half of reionisation at $z<8$.

Next, we estimate $\xi$, the exponent in \eref{eq:gam-xi}, from our simulation. To do this, we start by computing the Lyman-limit photon attenuation length as
\begin{equation}
  \label{eq:att-lam}
  \lambda = \left( n_{{\rm HI}}\sigma_{912} \right)^{-1},
\end{equation}
where $n_{{\rm HI}}$ is the number density of neutral hydrogen at a given position in the simulation, and $\sigma_{912}$ is the photoionisation cross-section at the Lyman-limit. Note that this is not equivalent to the mean free path of ionising photons defined in Eq.~(\ref{eq:mfp}).

Furthermore, \eref{eq:gam-xi} is derived under the assumption of photoionisation equilibrium in the reionised IGM, so to estimate $\xi$ we must also select regions where this condition holds in the simulation.  To do this, we require that
\begin{equation}
    \label{eq:pho-eq} 
    \mathcal{E} = \frac{x_{\rm HI}\Gamma}{n_{\rm H} (1+\chi_{\rm e})x_{\rm HII}^2 \alpha(T)} \approx 1,
\end{equation}
is satisfied, where $x_{\rm HI}$ is the neutral hydrogen fraction, $x_{\rm HII}=1-x_{\rm HI}$ is the ionised hydrogen fraction and $\chi_{\rm e}=0.079$ is a correction to account for electrons due to singly-ionised helium (assuming that helium is ionised once at the same time as hydrogen). In practice, when applying \eref{eq:pho-eq} we allow for some small deviation around equilibrium by applying a cut on $|{\rm log}_{10}\mathcal{E}|$.

\begin{figure}
    \centering
    \includegraphics[width=\linewidth]{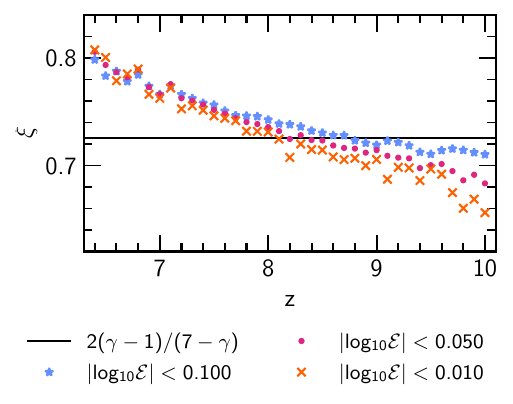}
    \vspace{-0.6cm}
    \caption{The value of the power-law slope, $\xi$, obtained by fitting \eref{eq:gam-xi} to the distribution of the attenuation length $\lambda=(n_{\rm HI} \sigma_{912})^{-1}$ and local photoionisation rate $\Gamma$ in the 40-2048 model as a function of redshift.  The coloured data points display the result when applying ionisation cuts to the simulation data that select for gas where ionisation equilibrium holds (see text for details).  The black horizonal line shows the analytically expected value of $\xi = 2(\gamma-1)/(7-\gamma)$ for $\gamma=2.6$.  \label{fig:xi-gam}}
\end{figure}

\fref{fig:xi-gam} shows the value of $\xi$ obtained by fitting \eref{eq:gam-xi} to the distribution of $\lambda=(n_{\rm HI} \sigma_{912})^{-1}$ against $\Gamma$ for (already reionised) gas at each redshift.  We find that, for reasonable values of $|{\rm log}_{10}\mathcal{E}|$, the choice of ionisation cut makes a small difference to the value of $\xi$ and no difference to the redshift evolution.  At $z\geq 6.3$, the typical value of $\xi$ we obtain directly from the simulation (data points in \fref{fig:xi-gam}) overlaps with the analytical scaling relation (solid black line) derived from the \mhr\ model for $\gamma=2.6$, although there is evidence that a $\xi$ that increases toward lower redshift may be more appropriate.  However, this agreement breaks at the point where the ionising radiation field is close to spatially uniform throughout the simulation box, and it is no longer possible to fit for the $\lambda$--$\Gamma$ correlation in the diffuse IGM. We therefore caution against extrapolation of the redshift dependence of $\xi$ to $z<6.3$.

Finally, returning to the \HI column density distribution function, $f(N_{\rm HI}, R)$, \citet{schaye2001} showed that the extent of overdense absorbers will typically be of order the local Jeans length.  Taking now the general expression for the proper Jeans scale (cf. \eref{eq:jeans-mean}), we have
\begin{equation}
  \label{eq:jeans-gen}
  \lambda_{\rm J} = \left( \frac{40 \pi^2 k_{\rm B} T}{9 \mu m_{\rm H} H^2_0 \Omega_{\rm m} (1+z)^{3} \Delta} \right)^{1/2}.
\end{equation} 
Dropping any constants and dependence on redshift or cosmology, we can write
\begin{equation}
  \label{eq:jeans-gen-prop}
  \lambda_{\rm J} \propto T^{1/2}\mu^{-1/2}\Delta^{-1/2}.
\end{equation} Using this relation in \eref{eq:pho-eq} we can express $x_{\rm HI}$, for highly ionised gas, as
\begin{equation}
  \label{eq:xhi-prop} x_{\rm HI} \propto T^{-0.72}\Delta\Gamma^{-1},
\end{equation} where we have used $\alpha(T) \propto T^{-0.72}$ \citep{VernerFerland1996,Bolton2022}. 
If the typical length of an absorber is $\lambda_{\rm J}$, then its typical column density is $N_{\rm HI}\approx \bar{n}_{\rm H} \Delta x_{\rm HI} \lambda_{\rm J}$, which we can express as
\begin{equation}
  \label{eq:nhi-col-prop} N_{\rm HI} \propto T^{-0.22}\mu^{-1/2}\Delta^{3/2}\Gamma^{-1}.
\end{equation} Next, following \citet{furlanetto2005}, we can relate the \HI column density distribution function to the underlying gas density distribution by
\begin{equation}
  \label{eq:fnhi-prop} f(N_{\rm HI}, R) \propto \frac{\xhi}{N_{\rm HI}} \Delta P(\Delta) \frac{\ud \Delta}{\ud N_{\rm HI}}.
\end{equation} We make one further assumption, namely that, for highly ionised gas $T$, $\mu$ and $\Gamma$ are all approximately constant, such that from \eref{eq:nhi-col-prop}

\begin{align}
  \label{eq:dnhi-ddel} \frac{\ud N_{\rm HI}}{\ud \Delta} &\simeq \left(\frac{\partial N_{\rm HI}}{\partial \Delta}\right)_{T, \mu, \Gamma}, \\ &\propto \Delta^{1/2}.
\end{align} This can be plugged into \eref{eq:fnhi-prop} and the other terms can be reduced to their relevant density dependences (i.e., $\xhi \propto \Delta$ and $N_{\rm HI}\propto\Delta^{3/2}$) to give
\begin{equation}
  \label{eq:fnhi-del-prop} f(N_{\rm HI}, R) \propto P(\Delta).
\end{equation} Finally, using the \mhr{} high density power-law fit to $P(\Delta)$, we can write $f(N_{\rm HI},R)\propto P(\Delta)\propto \Delta^{-\gamma}$ and hence
\begin{equation}
  \label{eq:fnhi-gam} f(N_{\rm HI},R) \propto N_{\rm HI}^{-2\gamma/3}.
\end{equation}
This result is plotted in Fig.~\ref{fig:column_density_threhold} for $\gamma=2.6$; we observe the slope is similar to the high column density tail of $f(N_{\rm HI},R)$ following reionisation.

% Don't change these lines
\bsp	% typesetting comment
\label{lastpage}
\end{document}